\documentclass[acmsmall,nonacm,preprint]{acmart}

\usepackage{import}
\usepackage{graphicx}
\usepackage{amsmath}
\usepackage{tabularx}
\usepackage{colortbl}
\usepackage{multirow}
\usepackage{color}
\usepackage{multicol}
\usepackage{booktabs}
\usepackage{array}
\usepackage{makecell}
\usepackage{caption}
\usepackage{subcaption}
\usepackage{url}
\usepackage{algpseudocode}
\usepackage{algorithm}
\usepackage{textcomp}
\usepackage{xcolor}
\usepackage{xspace}
\usepackage{comment}
\usepackage{enumitem}
\usepackage{xfrac}
\usepackage{adjustbox}
\usepackage[most]{tcolorbox}
\usepackage{graphicx}

\usepackage{pifont}

\usepackage{graphicx}
\usepackage{array}
\usepackage{lscape}
\usepackage{booktabs}
\usepackage{longtable}
\usepackage{tikz}
 
\newcommand*\halfcirc[1][0.55ex]{%
  \begin{tikzpicture}
  \draw[fill] (0,0)-- (90:#1) arc (90:270:#1) -- cycle ;
  \draw (0,0) circle (#1);
  \end{tikzpicture}}
 
\newcolumntype{N}{>{\centering\arraybackslash}m{.02in}}
\newcolumntype{G}{>{\centering\arraybackslash}m{0.5in}}

\usepackage{titlesec}
\usepackage{hyperref}
\usepackage{tabulary}

\titleclass{\subsubsubsection}{straight}[\subsection]

\newcounter{subsubsubsection}[subsubsection]
\renewcommand\thesubsubsubsection{\thesubsubsection.\arabic{subsubsubsection}}

\titleformat{\subsubsubsection}
  {\normalfont\normalsize\bfseries}{\thesubsubsubsection}{1em}{}
\titlespacing*{\subsubsubsection}
{0pt}{3.25ex plus 1ex minus .2ex}{1.5ex plus .2ex}

\makeatletter
\renewcommand\paragraph{\@startsection{paragraph}{5}{\z@}%
  {3.25ex \@plus1ex \@minus.2ex}%
  {-1em}%
  {\normalfont\normalsize\bfseries}}
\renewcommand\subparagraph{\@startsection{subparagraph}{6}{\parindent}%
  {3.25ex \@plus1ex \@minus .2ex}%
  {-1em}%
  {\normalfont\normalsize\bfseries}}
\def\toclevel@subsubsubsection{4}
\def\toclevel@paragraph{5}
\def\toclevel@paragraph{6}
\def\l@subsubsubsection{\@dottedtocline{4}{7em}{4em}}
\def\l@paragraph{\@dottedtocline{5}{10em}{5em}}
\def\l@subparagraph{\@dottedtocline{6}{14em}{6em}}
\makeatother

\AtBeginDocument{%
  \providecommand\BibTeX{{%
    \normalfont B\kern-0.5em{\scshape i\kern-0.25em b}\kern-0.8em\TeX}}}

\setcopyright{acmcopyright}
\copyrightyear{2018}
\acmYear{2018}
\acmDOI{XXXXXXX.XXXXXXX}

\acmJournal{JACM}
\acmVolume{37}
\acmNumber{4}
\acmArticle{111}
\acmMonth{8}

\begin{document}

\title{Evaluating the Security of Aircraft Systems}

\author{Edan Habler}
\email{habler@post.bgu.ac.il}
\author{Ron Bitton}
\email{ronbit@post.bgu.ac.il}
\author{Asaf Shabtai}
\email{shabtaia@bgu.ac.il}
\affiliation{%
  \institution{Ben-Gurion University of the Negev}
  \city{Beer-Sheva}
  \country{Israel}
}

\renewcommand{\shortauthors}{Habler, et al.}

\begin{abstract}
The sophistication and complexity of cyber attacks and the variety of targeted platforms have been growing in recent years.
Various adversaries are abusing an increasing range of platforms, e.g., enterprise platforms, mobile phones, PCs, transportation systems, and industrial control systems. 
In recent years, we have witnessed various cyber attacks on transportation systems, including attacks on ports, airports, and trains. 
It is only a matter of time before transportation systems become a more common target of cyber attackers.
Due to the enormous potential damage inherent in attacking vehicles carrying many passengers and the lack of security measures applied in traditional airborne systems, the vulnerability of aircraft systems is one of the most concerning topics in the vehicle security domain.
This paper provides a comprehensive review of aircraft systems and components and their various networks, emphasizing the cyber threats they are exposed to and the impact of a cyber attack on these components and networks and the essential capabilities of the aircraft.
In addition, we present a comprehensive and in-depth taxonomy that standardizes the knowledge and understanding of cyber security in the avionics field from an adversary's perspective. 
The taxonomy divides techniques into relevant categories (tactics) reflecting the various phases of the adversarial attack lifecycle and maps existing attacks according to the MITRE ATT\&CK methodology.
Furthermore, we analyze the security risks among the various systems according to the potential threat actors and categorize the threats based on STRIDE threat model.
Future work directions are presented as guidelines for industry and academia.
\end{abstract}




\maketitle

\section{\label{sec:intro}Introduction}
In light of the significant growth in the number of flights,\footnote{\scriptsize\url{https://www.statista.com/statistics/564769/airline-industry-number-of-flights/}} traditional air traffic management systems (ATMs) have difficulty providing relevant and reliable information on the state of an aircraft.
As a result, the aviation community is actively taking steps to increase flight traffic safety, capacity, and flexibility, as well as to reduce dependence on outdated infrastructure, by examining and improving national airspace systems. 
There are two main projects leading modernization efforts in the aviation industry: the NextGen\footnote{\scriptsize\url{https://www.faa.gov/nextgen/media/NextGenAnnualReport-FiscalYear2020.pdf}} and SESAR\footnote{\scriptsize\url{http://www.sesar.eu/}} projects, respectively led by the US Federal Aviation Administration (FAA) and the European Commission.
These projects are primarily aimed at creating new technologies and procedures in order to increase the capacity, accuracy, and reliability of air traffic control, and privacy and information security have not been prioritized.
Technological advances and modernization processes have been seen in many other areas, such as the transition to smart transportation, smartphones, and IoT components. At the same time, the attack surface accessible to cyber attacks is growing.
According to \textit{Positive Technology's} annual report, the first quarter of 2021 (Q1) showed a 17\% increase in cyber attacks over the same quarter in the previous year~\cite{PTsecurityRepport}. 
Moreover, according to Check Point's ransomware report~\cite{CheckPoint} between June 2020 and June 2021, the transportation industry witnessed a 186\% increase in weekly ransomware attacks.
Therefore, it is fair to assume that it is just a matter of time before we see significant cyber incidents in the avionics field.

As a case in point, in recent years, researchers in academia and industry have pointed out weaknesses in airborne systems' design and implementation and demonstrated how some of the core airborne systems could be tampered with simply by using commercial off-the-shelf (COTS) hardware and software; for instance, Costin~\textit{et al.}~\cite{costin2012ghost} simulated attacks on the ADS-B system using a COTS software-defined radio (SDR) transmitter, and Hugo~\textit{et al.}~\cite{teso2013aircraft} demonstrated how one can use a simple Android device to send radio signals and obtain access to the navigation controls of an aircraft by exploiting the aircraft communications addressing and reporting system (ACARS). 

To better classify the attacks associated with different threats, it is important to analyze the vulnerabilities of airborne systems, as well as the potential threat actors. 
Systematically categorizing adversaries' behavior will enable the identification of sensitive points in the various platforms and the stage of an ongoing attack, as well as preventive actions.

A few popular knowledge bases describe cyber adversary behavior and provide a common taxonomy for both offense and defense aimed at enterprise networks: FireEye's cyber kill chain,\footnote{https://www.fireeye.com/content/dam/fireeye-www/company/events/infosec/tech-track-summit-paris.pdf\label{FireEye}} Lockheed Martin's cyber kill chain, which is part of their intelligence-driven defense model for identification and prevention~\cite{hutchins2011intelligence}, and MITRE ATT\&CK~\cite{strom2018mitre}, which maintains a taxonomy for multiple platforms and networks (enterprise, mobile, and industrial control systems). 
However, no well-defined knowledge base aggregates all information regarding the threat components and provides a taxonomy for the different stages of attacks in the field of transportation and avionics.

In this paper, we address this gap by providing a broad overview of avionics systems, emphasizing the security aspects, and presenting a taxonomy of adversarial behavior associated with the communication, navigation, engagement, surveillance, and complementary systems and devices of aircraft. 
Following MITRE ATT\&CK's taxonomies, we have identified specific actions (techniques) and classified them under categories (tactics) to reflect the various phases of the adversarial attack lifecycle, and so we proposed an extension of the MITRE taxonomy adapted to aircraft.
Moreover, we present an ontology that defines the entities through which the various threats and actors that take part in the various attacks can be analyzed.

Since an aircraft is comprised of different communication systems and devices which have diverse roles in the aircraft's functionality, we examine aircraft according to their division into the domains defined by the Airlines Electronic Engineering Committee (AEEC) of ARINC~\cite{AEEC} in the \textit{ARINC Report 821}~\cite{ARINC821}: the passenger information \& entertainment domain (PIESD), passenger-owned devices domain (PODD), airline information services domain (AISD), and aircraft control domain (ACD).

Each domain contains its own components and processes, since an aircraft's various networks differentiate between different degrees of data sensitivity and logical responsibilities. Therefore, the adversaries' degree of influence depends on their presence. This division of domains can be used to isolate the weak points that can be used to spread between the domains.  
After performing this mapping and defining the avionics assets consisting of the various networks and components, we mapped the various attacks and threats that can be applied to them and analyzed them using the STRIDE~\cite{STRIDEMODEL} threat model, the attacker's capabilities, and the potential impact.

Table~\ref{tab:compare} compares our study to recent studies (\cite{elmarady2021studying,dave2022cyber,strohmeier2016perception,lykou2019aviation,shaikh2019review,strohmeier2020securing}) that analyze threats to various aviation systems.
To summarize, our contributions in this study are as follows:
First, we provide a broad overview of modern aircraft, covering their systems, domains, and networks, and emphasize their security gaps and known attack vectors and threats.
Second, we map the different attacks, and their required capabilities, targets, and threat categories.
Third, we provide a taxonomy that categorizes adversarial behavior targeting different attack surfaces and systems of aircraft; thus, we extended the MITRE taxonomy for the aviation field.
Finally, we present a test case of a cyber attack and demonstrate how it can be analyzed using the taxonomy.

\begin{table}[ht!]
\hspace*{-1.3cm}
\centering
\tiny
\begin{tabular}{
|m{0.04\textwidth}|NNN|NNNNNNNN|NNNN|NNNNN|NNNNNNNN| NNNN|m{0.01\textwidth}|G|}
\Xhline{2pt}

\multirow{3}{*}{\textbf{Paper}} &
\multicolumn{32}{c|}{\textbf{\underline{Assets Evaluated}}} &
\multirow{3}{0pt}{\makecell{\rotatebox{90}{\textbf{Taxonomy}}}} &
\multirow{3}{0pt}{\textbf{Security Model}}\\ 
    &
    \multicolumn{3}{c}{\textbf{\underline{Com.}}} &
    \multicolumn{8}{c}{\textbf{\underline{Navigation}}} &
    \multicolumn{4}{c}{\textbf{\underline{Surveillance}}} &
    \multicolumn{5}{c}{\textbf{\underline{Alerting}}} &
    \multicolumn{8}{c}{\textbf{\underline{Backend Avionic \& Network }}} &
    \multicolumn{4}{c|}{\textbf{\underline{Connectivity}}} & & \\
    & \rotatebox{90}{SATCOM}
    & \rotatebox{90}{CPDLC}
    & \rotatebox{90}{ACARS}
    & \rotatebox{90}{DME}
    & \rotatebox{90}{VOR}
    & \rotatebox{90}{NDB}
    & \rotatebox{90}{ILS}
    & \rotatebox{90}{GNSS}
    & \rotatebox{90}{ABAS}
    & \rotatebox{90}{SBAS}
    & \rotatebox{90}{GNAS}
    & \rotatebox{90}{PSR}
    & \rotatebox{90}{SSR}
    & \rotatebox{90}{MLAT}
    & \rotatebox{90}{ADS-B}
    & \rotatebox{90}{TCAS}
    & \rotatebox{90}{ACAS X}
    & \rotatebox{90}{Engine Alerting}
    & \rotatebox{90}{FIS-B}
    & \rotatebox{90}{TIS-B}
    & \rotatebox{90}{EFB}
    & \rotatebox{90}{TWLU / CWLU}
    & \rotatebox{90}{FMS}
    & \rotatebox{90}{EGM}
    & \rotatebox{90}{NIM}
    & \rotatebox{90}{CSS}
    & \rotatebox{90}{FDR}
    & \rotatebox{90}{CIS-MS}
    & \rotatebox{90}{ACD}
    & \rotatebox{90}{AISD}
    & \rotatebox{90}{PIESD}
    & \rotatebox{90}{External Networks} & & \\
\Xhline{2pt}

    \tiny{Our Study}
    & $\bullet$ %SATCOM 
    & $\bullet$ %CPDLC   
    & $\bullet$ %ACARS  
    & $\bullet$ %DME 
    & $\bullet$ %VOR 
    & $\bullet$%NDB
    & $\bullet$%ILS
    & $\bullet$%DNNSS
    & $\bullet$%ABAS
    & $\bullet$%SBAS
    & $\bullet$%GNAS
    & $\bullet$%PSR
    & $\bullet$%SSR
    & $\bullet$%MLAT
    & $\bullet$%ADS-B
    & $\bullet$%TCAS
    & $\bullet$%ACAS X
    & $\bullet$%Engine Alerting
    & $\bullet$%FIS-B
    & $\bullet$%TIS-B
    & $\bullet$%EFB
    & $\bullet$%TWLU 
    & $\bullet$%CWLU
    & $\bullet$%EGM
    & $\bullet$%NIM
    & $\bullet$%CSS
    & $\bullet$%FDR
    & $\bullet$%CIS-MS
    & $\bullet$%ACD
    & $\bullet$%AISD
    & $\bullet$%PIESD
    & $\bullet$%External Networks
    & $\bullet$%threat analysis
    &  STRIDE %taxonomy
    \\\hline

    \cite{dave2022cyber}
    & $\bullet$%SATCOM
    & $\bullet$%CPDLC
    & $\bullet$%ACARS
    & $\bullet$%DME
    & $\bullet$%VOR
    & $\circ$%NDB
    & $\bullet$%ILS
    & $\bullet$%DNNSS
    & $\circ$%ABAS
    & $\circ$%SBAS
    & $\circ$%GNAS
    & $\bullet$%PSR
    & $\bullet$%SSR
    & $\circ$%MLAT
    & $\bullet$%ADS-B
    & $\circ$%TCAS
    & $\circ$%ACAS X
    & $\circ$%Engine Alerting
    & $\circ$%FIS-B
    & $\circ$%TIS-B
    & $\circ$%EFB
    & $\circ$%TWLU
    & $\circ$%CWLU
    & $\circ$%EGM
    & $\circ$%NIM
    & $\circ$%CSS
    & $\circ$%FDR
    & $\circ$%CIS-MS
    & $\circ$%ACD
    & $\circ$%AISD
    & $\circ$%PIESD
    & $\circ$%External Networks
    & $\circ$%threat analysis
    & CIA%taxonomy
    \\\hline
    \cite{elmarady2021studying}
    & $\circ$ %SATCOM 
    & $\bullet$ %CPDLC   
    & $\bullet$ %ACARS  
    & $\bullet$ %DME 
    & $\bullet$%VOR 
    &  $\circ$%NDB
    & $\bullet$%ILS
    & $\bullet$%NSS
    & $\circ$%ABAS
    & $\circ$%SBAS
    & $\circ$%GNAS
    & $\bullet$%PSR
    & $\bullet$%SSR
    & $\bullet$%MLAT
    & $\bullet$%ADS-B
    & $\circ$%TCAS
    & $\circ$%ACAS X
    & $\circ$%Engine Alerting
    & $\circ$%FIS-B
    & $\circ$%TIS-B
    & $\circ$%EFB
    & $\circ$%TWLU
    & $\circ$%CWLU
    & $\circ$%EGM
    & $\circ$%NIM
    & $\circ$%CSS
    & $\circ$%FDR
    & $\circ$%CIS-MS
    & $\circ$%ACD
    & $\circ$%AISD
    & $\circ$%PIESD
    & $\circ$%External Networks
    & $\circ$%threat analysis
    & Likelihood \& impact %taxonomy
    \\\hline
    \cite{lykou2019aviation}
    & $\bullet$%SATCOM
    & $\circ$%CPDLC
    & $\circ$%ACARS
    & $\circ$%DME
    & $\circ$%VOR
    & $\circ$%NDB
    & $\circ$%ILS
    & $\bullet$%DNNSS
    & $\circ$%ABAS
    & $\circ$%SBAS
    & $\circ$%GNAS
    & $\bullet$%PSR
    & $\bullet$%SSR
    & $\circ$%MLAT
    & $\bullet$%ADS-B
    & $\bullet$%TCAS
    & $\circ$%ACAS X
    & $\circ$%Engine Alerting
    & $\bullet$%FIS-B
    & $\bullet$%TIS-B
    & $\circ$%EFB
    & $\circ$%TWLU
    & $\circ$%CWLU
    & $\circ$%EGM
    & $\circ$%NIM
    & $\circ$%CSS
    & $\circ$%FDR
    & $\circ$%CIS-MS
    & $\circ$%ACD
    & $\circ$%AISD
    & $\circ$%PIESD
    & $\circ$%External Networks
    & $\circ$%threat analysis
    & Categorized actors in relation to their resources and motivations%taxonomy
    \\\hline
    \cite{shaikh2019review}
    & $\circ$%SATCOM
    & $\bullet$%CPDLC
    & $\bullet$%ACARS
    & $\circ$%DME
    & $\circ$%VOR
    & $\circ$%NDB
    & $\circ$%ILS
    & $\bullet$%DNNSS
    & $\circ$%ABAS
    & $\circ$%SBAS
    & $\circ$%GNAS
    & $\circ$%PSR
    & $\circ$%SSR
    & $\circ$%MLAT
    & $\bullet$%ADS-B
    & $\circ$%TCAS
    & $\circ$%ACAS X
    & $\circ$%Engine Alerting
    & $\circ$%FIS-B
    & $\circ$%TIS-B
    & $\circ$%EFB
    & $\circ$%TWLU
    & $\circ$%CWLU
    & $\circ$%EGM
    & $\circ$%NIM
    & $\circ$%CSS
    & $\circ$%FDR
    & $\circ$%CIS-MS
    & $\circ$%ACD
    & $\circ$%AISD
    & $\circ$%PIESD
    & $\circ$%External Networks
    & $\circ$%threat analysis
    & Discussed of security challenges %taxonomy
    \\\hline
    \cite{strohmeier2020securing}
    & $\circ$%SATCOM
    & $\bullet$%CPDLC
    & $\bullet$%ACARS
    & $\circ$%DME
    & $\circ$%VOR
    & $\circ$%NDB
    & $\circ$%ILS
    & $\circ$%DNNSS
    & $\circ$%ABAS
    & $\circ$%SBAS
    & $\circ$%GNAS
    & $\circ$%PSR
    & $\bullet$%SSR
    & $\bullet$%MLAT
    & $\bullet$%ADS-B
    & $\bullet$%TCAS
    & $\circ$%ACAS X
    & $\circ$%Engine Alerting
    & $\circ$%FIS-B
    & $\circ$%TIS-B
    & $\circ$%EFB
    & $\circ$%TWLU
    & $\circ$%CWLU
    & $\circ$%EGM
    & $\circ$%NIM
    & $\circ$%CSS
    & $\circ$%FDR
    & $\circ$%CIS-MS
    & $\circ$%ACD
    & $\circ$%AISD
    & $\circ$%PIESD
    & $\circ$%External Networks
    & $\circ$%threat analysis
    & Examined realistic events and accidents that have occured%taxonomy
    \\\hline
    \cite{strohmeier2016perception}
    & $\circ$%SATCOM
    & $\bullet$%CPDLC
    & $\bullet$%ACARS
    & $\bullet$%DME
    & $\bullet$%VOR
    & $\bullet$%NDB
    & $\bullet$%ILS
    & $\circ$%DNNSS
    & $\circ$%ABAS
    & $\circ$%SBAS
    & $\circ$%GNAS
    & $\bullet$%PSR
    & $\bullet$%SSR
    & $\bullet$%MLAT
    & $\bullet$%ADS-B
    & $\bullet$%TCAS
    & $\circ$%ACAS X
    & $\circ$%Engine Alerting
    & $\bullet$%FIS-B
    & $\bullet$%TIS-B
    & $\circ$%EFB
    & $\circ$%TWLU
    & $\circ$%CWLU
    & $\circ$%EGM
    & $\circ$%NIM
    & $\circ$%CSS
    & $\circ$%FDR
    & $\circ$%CIS-MS
    & $\circ$%ACD
    & $\circ$%AISD
    & $\circ$%PIESD
    & $\circ$%External Networks
    & $\circ$%threat analysis
    & Operational impact%taxonomy
    \\\Xhline{2pt}
    
\end{tabular}

\caption{Comparison of the proposed threat analysis with existing aviation security surveys.}
\label{tab:compare}
\end{table}

\section{Threat Analysis\label{sec:ontology}}
\begin{figure}[ht!]
\centering
\includegraphics[width=0.8\textwidth]{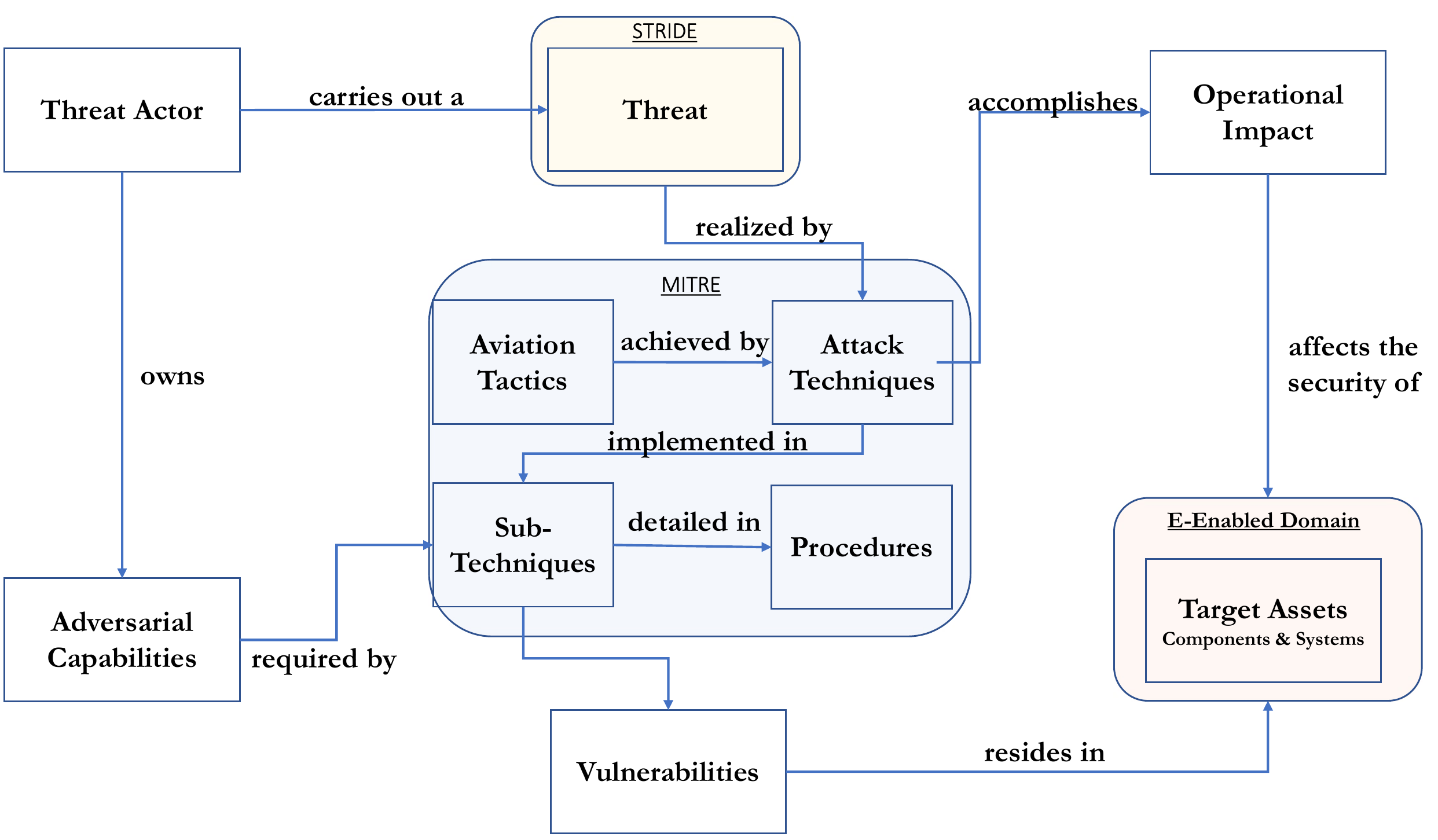}  
\caption{Threat analysis ontology.}
\label{fig:threat_ontology}
\end{figure}

Our security analysis is performed according to the ontology presented in Figure~\ref{fig:threat_ontology}, which is based on the NIST ontology for evaluating enterprise security risk.
The threat analysis ontology includes the following entities:
\begin{itemize}[leftmargin=*]
    \item \textbf{Threat Actor.} An individual, group, or state responsible for an event or incident that impacts, or has the potential to impact, the security or safety of an aircraft system. 
    Threat actors possess different adversarial capabilities and therefore may execute different attack techniques.
    \item \textbf{Threat.} A threat represents a potential violation of a security property, such as integrity, confidentiality, or availability. In a threat, an attacker exploits a vulnerability in order to perform an attack. In this paper, we opt to use STRIDE threat model to identifying security threats.
    \item  \textbf{Adversarial Capabilities.} Refers to the capabilities that are available to the threat actor. Within the context of aviation, we distinguish between access, positional, knowledge and material capabilities. 
    \item \textbf{Vulnerability.} A characteristic of an asset or technology that makes it prone to an attack.
    \item \textbf{Operational Impact.} The operative effect for which an attack was executed.
    \item \textbf{E-Enabled Domains.} A division of different communication systems and devices the aircraft is comprised of, which have diverse roles in an aircraft’s functionality.
    \item \textbf{Target Assets.} The main data, systems, components, processes, and services that comprise an aircraft system and should be protected.
    \item \textbf{Aviation Tactics.} Refers to the adversary's tactical goal: the reason for performing an action. 
    \item \textbf{Attack Technique.} An act or method that can be used by threat actors to realize threats.
    \item \textbf{Sub-Techniques.} A more specific description of the adversarial behavior used to achieve a goal. In this work we opt to consider this as a description of a concrete attack which was implemented by at least one party.
    \item \textbf{Procedures.} The specific details of how an adversary carries out a technique to achieve a tactic. Since it is not yet possible to refer to specific groups which executed attacks, the procedures in this work are references to academic works or practical threats presented in the industry.
\end{itemize}

In order to analyze the threat model and define a taxonomy for attacks in the aviation field, we performed the following steps:

First, we performed a broad review of the various avionics systems, communication networks, and the back-end components that comprise the aircraft's complete assembly.
This review was carried out by examining the \emph{e-Enabled domains} in order to identify the various \emph{target assets} (Sections~\ref{sec:systemreview} - \ref{sec:com}).
Second, for each asset we mapped the \emph{threats} and concrete attacks that were carried out on it (detailed in Section~\ref{sec:cybersecurity}) in order to identify its inherent \emph{vulnerabilities}.\\
Third, we analyzed the various attacks using the \emph{STRIDE threat model} and mapped the \emph{threat actors}(Section~\ref{sec:threatactor}), while identifying the various \emph{adversarial capabilities} that are required in order to implement the attacks carried out (Section~\ref{sec:Adversaries}).\\
Fourth, we analyzed the concrete attacks carried out in academia and industry and distilled from them the \emph{sub-techniques} that were used to carry out the various attacks, summarized in Table~\ref{tab:CIATriangle}).\\
Finally, through these mappings of the potential threat actors, their capabilities, and the concrete attacks that were carried out in industry and academia, we defined taxonomy for the aviation field, organized as an extension of the MITRE framework (Section~\ref{sec:taxonomy}). 
Our proposed taxonomy is defined by abstracting the different concrete attacks and \emph{techniques} used by the attackers and defining them as a means to achieve a goal as part of a multi-stage attack. 
A demonstration for modeling the proposed taxonomy framework is presented in Section~\ref{sec:usecases}.

\section{\label{sec:systemreview}The Target Assets of an Aircraft System}

In the last decade, the architecture of avionics and information systems in aircraft has evolved and developed to enable real-time data links between aircraft and the ground for information sharing, i.e., passing critical control, maintenance, navigation, and operations data. 
There was also a need to reduce the weight of computer and network infrastructure on  aircraft to reduce the cost of fuel.
This was achieved by consolidating a number of software systems on integrated modular avionics (IMA) computing modules capable of supporting numerous applications and transitioning  to Ethernet-based protocols and networks (and more specifically to Avionics Full-Duplex Switched Ethernet, defined in the ARINC-664 report~\cite{ARINC664}).

\begin{figure}[h!]
    \centering
\includegraphics[width=0.75\linewidth, keepaspectratio]{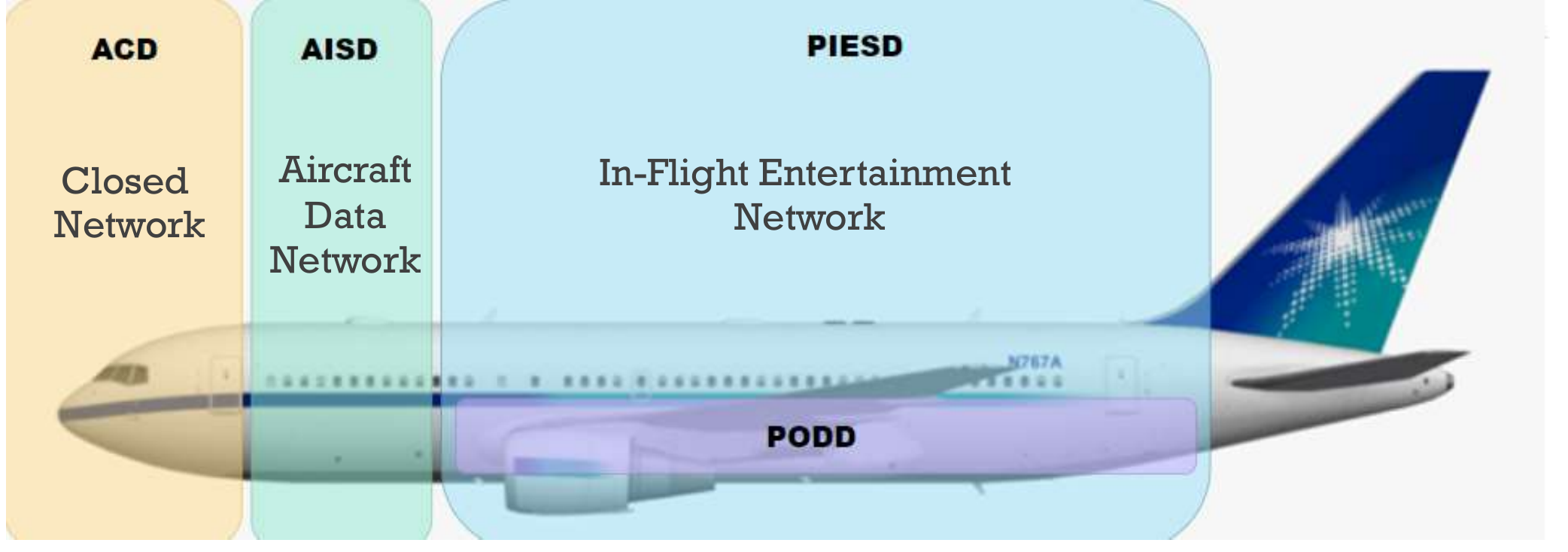}
    \caption{e-Enabled aircraft domains.}
    \label{fig:domains}
\end{figure}

Aircraft systems like passenger engagement systems and critical navigation systems have different roles, and levels of importance and sensitivity.
Therefore, the target assets of aircraft systems are divided into three e-Enabled domains (see Figure~\ref{fig:domains}):
the Aircraft Control Domain (ACD), Airline Information Services Domain (AISD), and Passengers (PIESD and PODD).

\begin{figure}[ht!]
    \centering
    \includegraphics[width=0.75\linewidth]{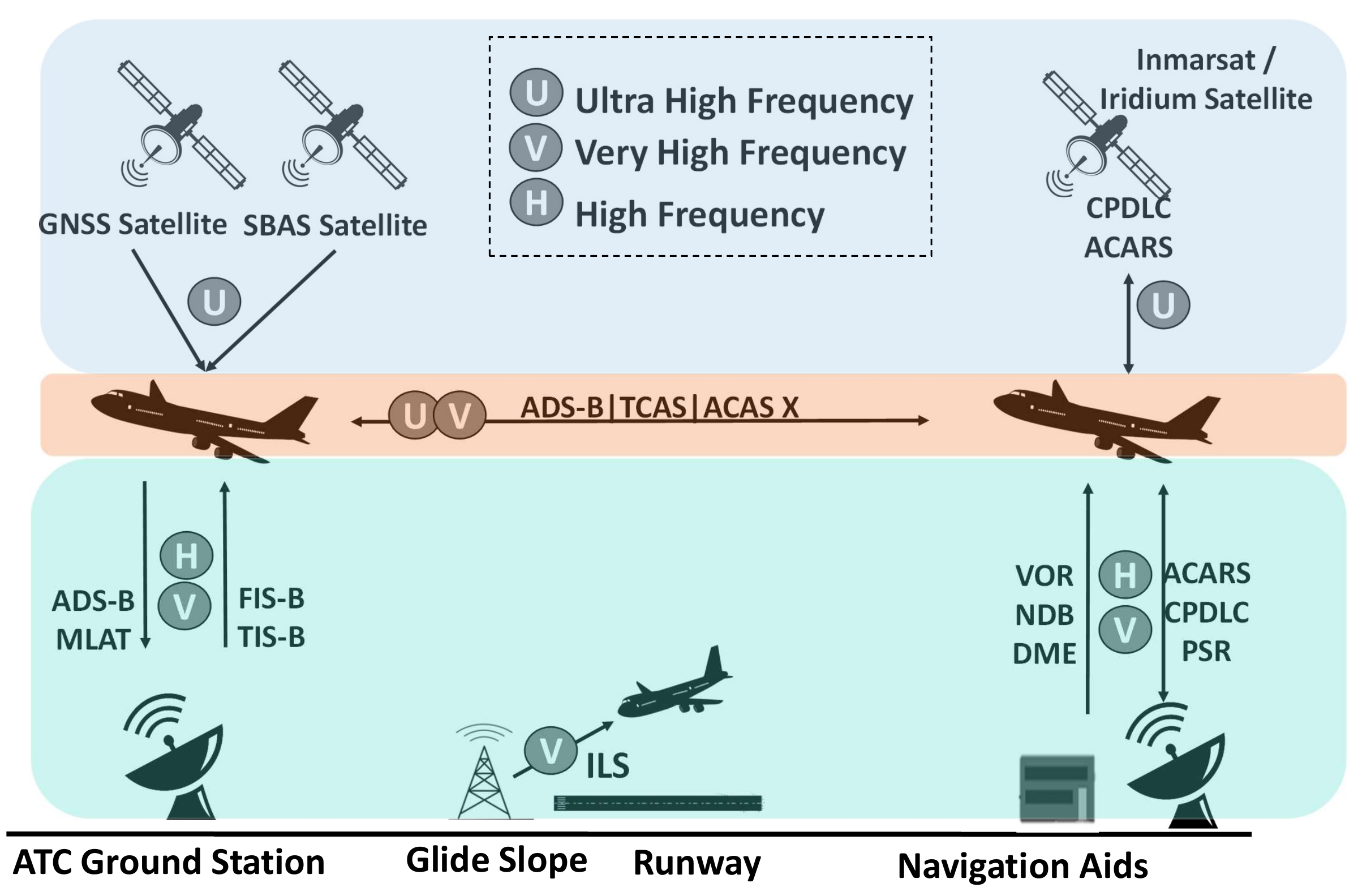}
    \caption{Aviation systems and technology infrastructure, positioning and range of operation.}
    \label{fig:ALLSYSTEMS}
\end{figure}

\begin{figure}[ht!]
    \centering
    \includegraphics[width=0.95\linewidth, keepaspectratio]{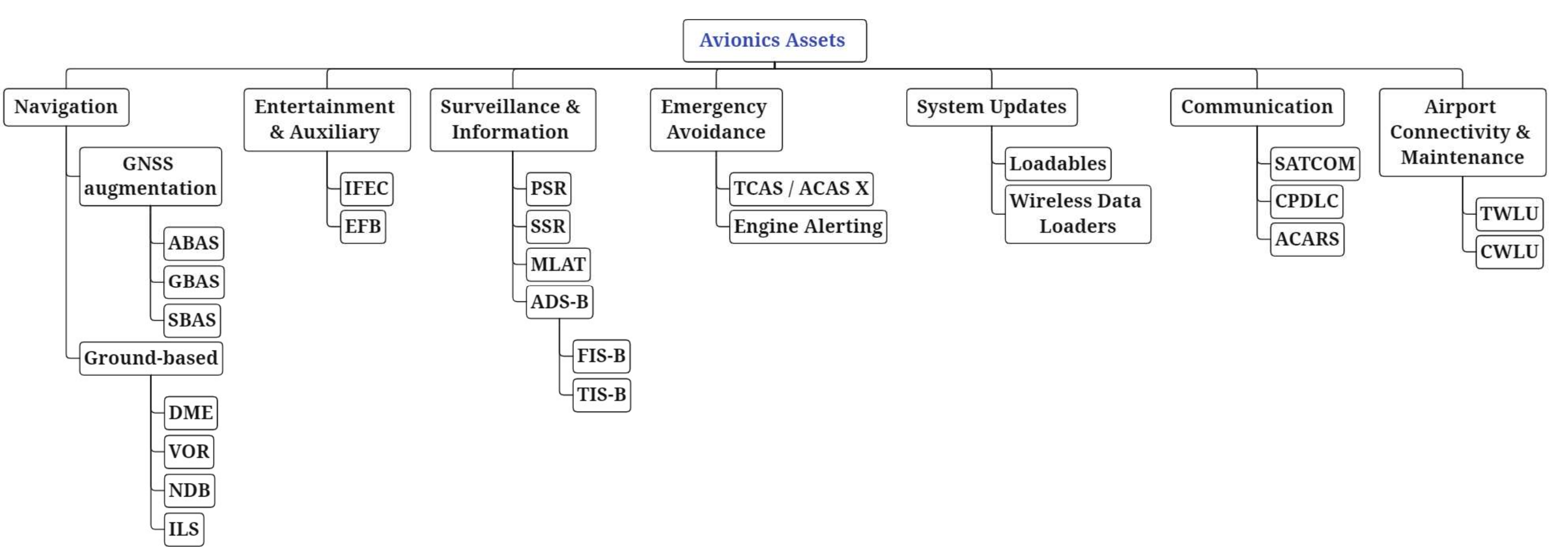}
    \caption{Categorization of Air/Ground avionics systems - Navigation systems, Entertainment systems, Surveillance systems, Information systems, Communication systems and Maintenance systems.}
    \label{fig:avionicssystem}
\end{figure}

In this section, we briefly describe each of the aircraft e-Enabled domains, analyze the various target assets which are part of each domain, and analyze the connectivity between different domains. 
Figure~\ref{fig:ALLSYSTEMS} illustrates the various systems that communicate with the aircraft, their location, and means of communication; Figure \ref{fig:avionicssystem} illustrates the different avionics assets, categorized according to their purpose and infrastructure.

\subsection{Aircraft Control Domain (ACD)}
The ACD consists of the systems and networks responsible for the aircraft's safe operation, e.g., air traffic control (ATC) and aircraft operational control (AOC) communications. 
Therefore, the ACD has the most stringent security requirements.
The ACD contains two subdomains: the \textit{Cabin-Core} and the \textit{Flight and Embedded Control System}.
The \textit{Cabin-Core} is designed to provide the services required to operate the cabin components, such as public address systems, smoke detectors, and air conditioning. 
The \textit{Flight and Embedded Control System} is designed to allow the pilot to control the aircraft from the flight deck.

To communicate with on-ground services, the aircraft use different components such as optical diodes and electronic gateway modules.
Sometimes, these components are the same as the components that serve the passenger domains; thus the separation between domains is not absolute.

There are a large number of critical systems used by the aircraft during flight. 
We mapped the main systems according to the nature of their function: 
surveillance, communication, emergency avoidance, system updates, navigation, entertainment, and auxiliary (this mapping is illustrated in Figure~\ref{fig:avionicssystem}).

\subsubsection{Surveillance and Information Systems} 
\begin{flushleft}
The surveillance systems are responsible for proactive, comprehensive monitoring of the aircraft location, while the information systems are responsible for supporting the aircraft operators and providing aeronautical information.
\end{flushleft}

\begin{itemize}[leftmargin=*]
    \item{\textbf{ADS-B system:}}
    Automatic Dependent Surveillance–Broadcast (ADS-B) is a satellite-based, `radar-like' system designed to continuously derive the aircraft position from the global navigation satellite (GPS) system. 
    ADS-B provides the aircraft's position and velocity with high accuracy, providing a clearer picture of the air traffic than traditional radar systems. 
    ADS-B includes two separate systems: ADS-B In and ADS-B Out.
    The ADS-B In system allows an aircraft to receive and display messages transmitted by other aircraft within the receiving range. 
    The ADS-B Out system allows an aircraft to continuously generate and broadcast messages over an unencrypted L-band range of frequencies.
    There are several types of certified ADS-B data links, but the preferred international link operates at 1090 MHz.
    Each ADS-B report may contain the following aircraft attributes: \textit{ICAO code (unique identifier), time, longitude, latitude, altitude, and velocity.}
    
    \item{\textbf{FIS-B:}}
    Flight Information System Broadcast (FIS-B) is a service that operates along with the ADS-B system to allow aircraft operators to obtain aeronautical information (e.g., temporary flight restrictions and weather reports). 
    FIS-B data is transmitted from ground stations to the aircraft's ADS-B receiver, which is responsible for processing the data and displaying a graphical national weather service data and flight restrictions on cockpit displays.
    
    \item{\textbf{TIS-B:}}
    Traffic Information Services-Broadcast (TIS-B) is another service that operates with the ADS-B system.
    TIS-B is a surveillance service which uses the ADS-B ground stations and radar data to transmit aircraft position data to aircraft cockpit displays. 
    Using TIS-B, the air traffic controller can share information regarding nearby aircraft with the pilot.
    
    \item{\textbf{PSR:}}
    Primary Surveillance Radar (PSR) is a surveillance radar system that does not require any onboard equipment to locate aircraft. 
    PSR uses a radar antenna to emit a radio wave pulse. 
    When directed at an aircraft the wave is reflected, and the resulted energy is returned back to the PSR antenna. 
    By analyzing the reflected pulse, the system can infer the range and bearing of the aircraft with respect to the antenna position.
    Since the PSR does not consist of onboard components, it does not represent an avionics asset for the purposes of our research and therefore we will not expand on it further in this paper.
        
    \item{\textbf{SSR:}}
    Secondary Surveillance Radar (SSR) is a surveillance radar system which uses transmitters and transponders as interrogators.
    The SSR radar antenna transmits a pulse which is received by onboard transponder. 
    The transponder returns a reply that contains information regarding the aircraft state (e.g., identity code, aircraft's altitude).
    
    \item{\textbf{MLAT:}}
    Multilateration (MLAT) is a technology that is used for navigation and surveillance.
    MLAT uses a methodology of analyzing a signal's time difference of arrival (TDOA) using multiple sensors in fixed locations, to infer the aircraft transmitter location.
    In aviation, MLAT is used by many stakeholders (e.g., live flight trackers) to track aircraft location. 
\end{itemize}

\subsubsection{Emergency Avoidance Systems}
\begin{flushleft}
Emergency avoidance systems are designed to issue an alert when there is a critical failure (e.g., engine failure), increase cockpit awareness of nearby aircraft, and serve as the last defense against mid-air collisions.
\end{flushleft}

\begin{itemize}[leftmargin=*]
    \item \textbf{TCAS/ACAS X:}
    The traffic collision avoidance system (TCAS) is designed to issue alerts and prevent mid-air collisions. The TCAS uses an onboard surveillance system to interrogate the airspace around an aircraft for other aircraft equipped with an active corresponding transponder (Mode-S transponder). The transponder is used to transmit signals indicating the aircraft's position, altitude, and vertical speed.
    The TCAS consists of two antennas, one of which is loated on top of the fuselage, and other of which is located on the bottom of the fuselage; then the aircraft monitored appears on the navigation display.
    By analyzing the replies from nearby aircraft, the TCAS can predict a potential collision, raise an alert regarding a potential intruder (a nearby aircraft), and request a resolution advisory (manoeuvre instruction to prevent collision) for both of the aircraft. 
    In the last decade, the FAA has funded research aimed at developing a modern approach to collision avoidance---a collision avoidance system known as ACAS X, which will use dynamic programming and provide more accurate alerts; in addition, ACAS X aims to support aircraft equipped with just passive surveillance mechanisms.
    \item \textbf {Engine alerting system:}
    The engine alerting system enables the flight crew to visualize the engine parameters and faults.
    There are two commonly used systems for engine alerting:
    The Engine-Indicating and Crew-Alerting System (EICAS) is an implementation of an engine alerting system.
    The EICAS consists of two monitors, two computers, and a display select panel.
    The monitors are used to display engine status and maintenance information, the display select panel enables the pilot to determine which of the two computers provides the engine information to the monitors, as one of the computers provides data and the other one serves as a standby.
    When an engine failure occurs, the system alerts the pilot, and the parameters of the event are recorded so they can be analyzed afterward by relevant experts.
    While the EICAS is mainly deployed on Boeing aircraft, the Electronic Centralized Aircraft Monitor (ECAM) is the corresponding system deployed on Airbus aircraft. 
    The main difference between the EICAS and ECAM is that ECAM lists the actions required to deal with a failure.
\end{itemize}

\subsubsection{Navigation Systems}
\begin{flushleft}
The navigation systems are designed to assist navigation throughout all phases of the flight, from takeoff to landing.
These systems operate in different terrain conditions and at different distances from the ground.
\end{flushleft}

\begin{itemize}[leftmargin=*]
    \item \textbf{GNSS and augmentation systems:}     The global navigation satellite system (GNSS) refers to a constellation of satellites providing signals from space that transmit positioning and timing data to GNSS receivers. The receivers then use this data to determine the position and velocity of the aircraft.
    In the aviation field, GNSS is used as a necessary utility in different systems, e.g., the ADS-B system uses GNSS to provide aircraft positions to ATC.
    \begin{enumerate}
        \item \textbf{ABAS} is an augmentation system that integrates the information obtained from the GNSS with information available on board the aircraft.
        \item \textbf{GBAS} is an augmentation system that that provides integrity monitoring using data obtained from ground sensors.
        \item \textbf{SBAS} is an augmentation system that improves the integrity, accuracy, and reliability of the GPS signal, using a number of geostationary satellites that cover vast areas.
    \end{enumerate}
    \item \textbf{Ground-based navigation systems:}
        There are four main kinds of ground-based navigation systems:
        \begin{enumerate}
        \item \textbf{ILS - }The instrument landing system (ILS) is a ground-based radio navigation system that provides short-range precision guidance to an aircraft approaching a runway with complex visibility conditions affected by light or weather.
        The ILS uses very high frequency (VHF) electromagnetic waves to provide horizontal guidance and ultra high frequency (UHF) electromagnetic waves to provide vertical and range guidance.
        The main components of ILS are the localizer, glide slope, and marker beacons.
        \item \textbf{VOR system - } The VHF omnidirectional radio range (VOR) is a ground-based navigation aid system operating in the VHF band. An aircraft equipped with a VOR receiver can determine its clockwise bearing from magnetic north, with reference to the ground station, by transmitting VHF navigation signals at radial angles
        \item \textbf{DME system - } Distance measuring equipment (DME) is a form of ground-based radio navigation aid that uses interrogation to compute the distance between an aircraft and ground DME equipment.
        The aircraft transmits a signal which is returned by the DME ground equipment after a fixed delay. The aircraft's distance from the ground equipment can be measured based on the delay of the returned signal perceived by the aircraft's DME equipment.
        \item \textbf{NDB system -} The non-directional radio beacon (NDB) is a ground-based navigation aid system which, in contrast to the VOR system, does not include inherent directional information.
        The NDB signals follow the curvature of the Earth, and thus they can be received from much greater distances at lower altitudes. NDB requires knowledge of the aircraft's exact heading to provide high accuracy, while VOR does not.
    \end{enumerate}
\end{itemize}

\subsection{Airline Information Services Domain (AISD)}
The AISD provides different types of services for non-essential/third-party applications, e.g., computing power, data storage, and routing. 
Independent aircraft applications such as avionics, in-flight entertainment, flight crew, and flight attendant applications use the AISD for connectivity purposes.
The AISD contains two subdomains: the \textit{Administrative} and \textit{Passenger Support} subdomains.
The \textit{Administrative} subdomain is designed to provide the flight deck and cabin with operational and administrative information. 
The \textit{Passenger Support} subdomain is designed to provide information to the passengers.

\begin{itemize}[leftmargin=*]
    \item \textbf{EFB: } The EFB (electronic flight bag) server is a highly customizes and flexible component of the aircraft, which is used for information management.
    The EFB is connected to most of the aircraft's avionic systems and sensors via dedicated interfaces (e.g., ARINC 615 and ARINC 429).
    Traditionally, Class 3 EFB systems are installed as aircraft equipment that includes an EFB server and a dedicated multi-function display. 
    In recent years, pilots have started using portable and commercial tablets  provided by the airlines as an extension to the EFB (Considered as Class 2 EFB systems).\footnote{https://www.neowin.net/news/delta-airlines-to-equip-pilots-with-surface-2-tablets/} 
    Class 2 EFB systems usually contain applications that are complementary to the services provided by the EFB, such as calculation of the take-off data.
\end{itemize}

\subsection{Passenger Domain (PIESD)}
The passenger domain can be divided into two subdomains:
\textit{Passenger Information \& Entertainment Services Domain (PIESD)} and \textit{Passenger Owned Devices Domain (PODD)}.
The PIESD is designed to serve the passengers, providing them with Internet and entertainment services. 
In addition to traditional entertainment systems, this domain allows access to wireless networks, links to passengers' physical devices, and seat adjustments.
The PIESD also connects passengers with the flight information system.
The PODD consists of external devices that passengers bring on board. 
To connect these external devices to the aircraft system, a passenger has to go through the PIESD.

\begin{itemize}[leftmargin=*]
    \item \textbf{IFEC system: }
    The in-flight entertainment and communication (IFEC) system refers to the entertainment applications available to passengers during flight (e.g., TV, audio, Wi-Fi, maps, and games).
    The IFEC includes content communication systems from external providers that are designed to enable telephony, satellite, and Internet services. These systems usually include display screens, computers with Linux/Windows/Android operating systems, and hosting/storage servers.
\end{itemize}

\section{Domain Connectivity}
The aircraft networks include overlapping components and networks that may allow attackers to move between the domains and components. Therefore, to defend against an attacker's trying to spread within the aircraft network, there is great value in locating potential areas that an attacker can exploit for lateral movement.

\begin{figure}[ht!]
    \centering
    \includegraphics[width=0.9\linewidth, keepaspectratio]{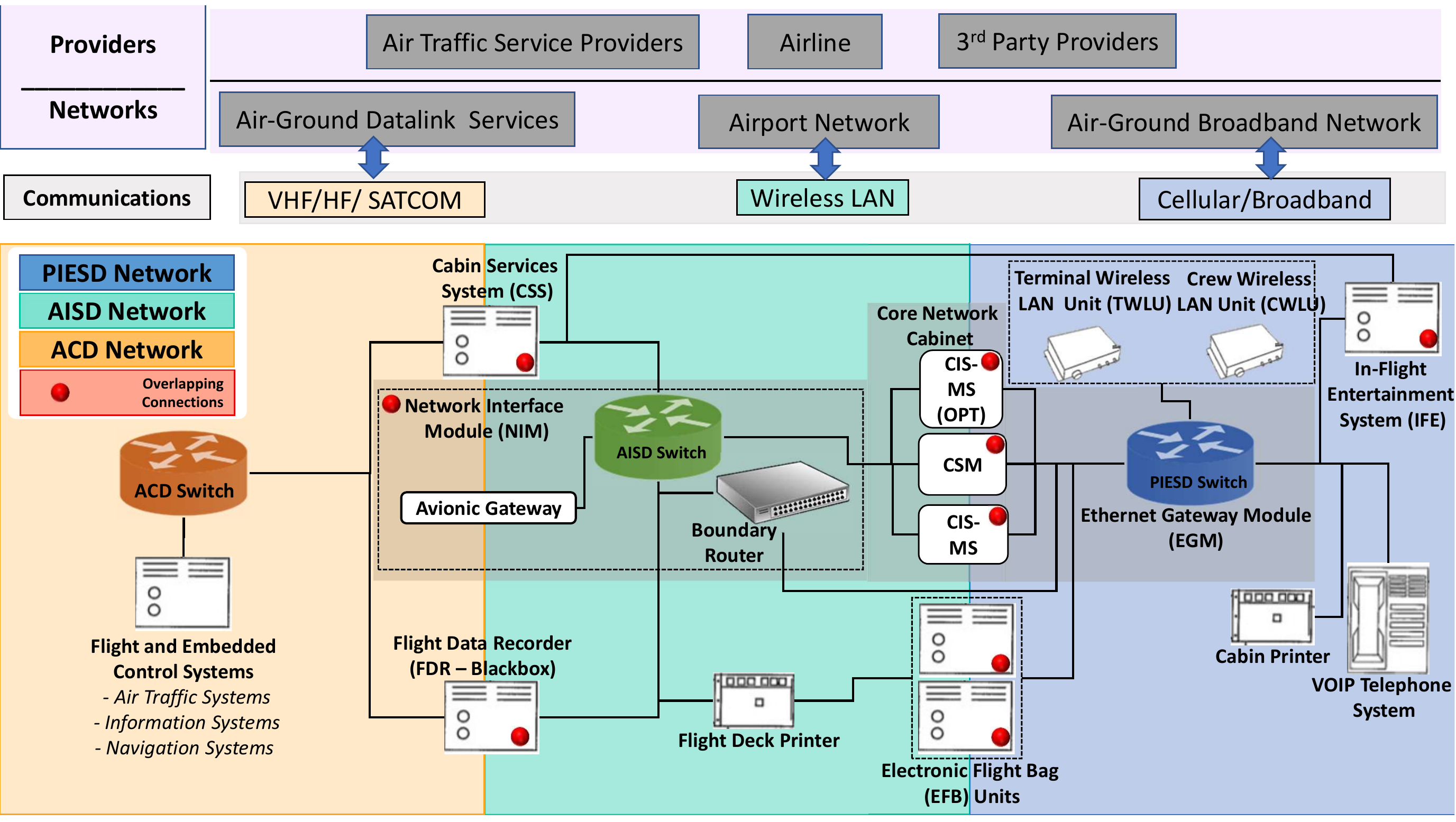}
    \caption{E-enabled implementation - system and component connectivity.
    The figure highlights the backend components that comprise the various domains; overlapping components are marked with a red circle.}
    \label{fig:eenabled}
\end{figure}

Figure~\ref{fig:eenabled} illustrates the
division into the different domains, their backend components, and the components that link them. A red dot marks the overlapping components between the grids.
Moreover, the figure contains the means of communication between the different domains to the providers.

\subsection{PIESD and AISD Connectivity}\label{PIESDAISD}
The core network cabinet is responsible
for data segregation between the PIESD, AISD, and ACD. The core network cabinet consists of several components, i.e., Ethernet gateway module (EGM), the controller server module (CSM), and the crew information system/maintenance system (CIS-MS) file server module (FSM).
These components are intended to provide cabin services, but they must be able to access the AISD for this purpose.

\begin{itemize}[leftmargin=*]
  \item The EGM includes an Ethernet switch and router for managing the PIESD and AISD connectivity.
  \item The CSM utilizes a dual connection, providing network management services (e.g., DHCP and DNS services) to the PIESD and fault reporting across the AISD and therefore to the ACD which hosts the maintenance system.
  \item The CIS-MS FSM is dually connected to provide the ACD/AISD/PIESD systems' data load services.
  These services include file transfer, data retention, wireless device control, and communication services.
\end{itemize}
In addition to the core network cabinet, the EFB also connects the networks. 
The EFB is an electronic information management device that helps flight crews perform flight management tasks more easily and efficiently. 
The EFB is connected to the PIESD and AISD via the AISD switch and the EGM. 
This interconnection allows, for example, problem reports to be offloaded from the aircraft in the event of wireless/broadband satellite, AISD, or PIESD failure, using the e-logbook application hosted on the EFB.
\subsection{AISD and ACD Connectivity}\label{AISDACD}
The network interface module (NIM) includes an avionics gateway which provides network address and protocol translation to connect the AISD to the more secure avionics in the ACD through the ACD switch.
To improve the fault isolation capability of the maintenance system, in case of a failure of the NIM boundary router and/or the EGM, the following aircraft systems may also be connected to the ACD switch:
The flight data recorder (FDR) and cabin services system (CSS).
\begin{itemize}[leftmargin=*]
  \item The FDR is an optional server that records high-value data that must be available at all times.
  \item The CSS has dual connections: the connectivity to the in-flight entertainment (IFE) system enables information to be presented to and played by passengers, while the connectivity to the ACD and  AISD switches enables the transfer of audio between the pilot and the cabin.
\end{itemize}
\subsection{External Connectivity}\label{EXTERNALCONNECTION}
Several external connections allow remote access to the aircraft network for the benefit of synchronization, updating information, and maintenance.

\begin{enumerate}[leftmargin=*]
  \item Maintenance laptop connectivity is possible using the connectivity and crew wireless LAN unit thus enabling external components to access the aircraft network.
  \item External flight planning systems can communicate with the aircraft crew via ACARS to update navigation plans.
  \item External terminals provided by trusted parties can be used for wireless communication with data-loaders that are used for software updates.
  \item The aircraft terminal wireless LAN unit (TWLU) is used for airport gatelink connectivity, while the connectivity and crew wireless LAN unit (CWLU) is used for maintenance laptop connectivity.
  \item the airport wireless network can be used to access devices on the same local area network, and these components (such as an EFB tablet or passengers' private devices) can have an impact on the aircraft. 
 \end{enumerate}

\begin{itemize}[leftmargin=*]
    \item \textbf{Loadable system:}
    A loadable system consists of a loadable replaceable unit (LRU), which is a modular component that is designed to be replaced quickly, and software that can be transferred to an LRU to modify system functionality. 
    Software updates are used for a number of purposes:
    \begin{enumerate}
    \item {\textit{OPS and OPC:}} update the operational program software/configuration (the operating system data and configurations of the LRUs).
    \item {\textit{Databases:}} update several databases, e.g., the engine data, flight management computer, and flight plans.
    \item {\textit{AMI (airline modifiable information):}} The AMI defines software generated by the operator to customize system operations, e.g., customization of the control display unit screens that are displayed to the flight crew.
    \end{enumerate}
    The loadable software updates can be transferred to the aircraft using a physical disk or wireless data loaders (e.g., Teledyne technologies' LoadStar server). 
\end{itemize}

\section{\label{sec:com}Communication Systems}
\subsection{SATCOM}
The satellite communication system (SATCOM) is used for reliable data and voice communication. 
The SATCOM serves as a data link for different uses, such as ADS-B, controller pilot data link communications (CPDLC), and the ACARS.
The SATCOM is comprised of the following components:
\begin{enumerate}[leftmargin=*]
\item {\textit{Satellite data unit (SDU):}} the SDU allows air and ground communication via a satellite network. The SDU uses a radio frequency unit to connect with a satellite. 
\item {\textit{Low and high gain antennas:}} The antennas contain an integrated beam steering unit and receive command information directly from the SDU.
\end{enumerate}
\subsection{CPDLC/FANS-1/ATN}
The CPDLC provides a means of communication between air traffic controllers and pilots over a data link system. The CPDLC data link is used to transmit non-urgent strategic messages to an aircraft and serves as an alternative to voice communications.
There are two popular implementations of the CPDLC data link system: the future air navigation system (FANS-1) and the aeronautical telecommunication network (ATN) system.
FANS-1 is an ACARS-based service that mainly uses Inmarsat satellite communications services, and the ATN is based on a VHF data link and is mainly operated by ARINC and SITA.  
\subsection{ACARS}
The aircraft communications addressing and reporting system (ACARS) is a digital data link system that enables the transmission of short messages between aircraft and ground stations through a network of transceivers.
ACARS messages are used to communicate with ATC and the base operational office.
This system is most often used for transmitting departure information, weather information, aeronautical operational control information, and OOOI events ([gate] Out, [wheels] Off, [wheels] On and [gate] In), which are automatically collected and represent the flight phase and related information (e.g., amount of fuel).
The main components of the ACARS are:

\begin{enumerate}[leftmargin=*]
\item {\textit{Onboard equipment:}} The onboard equipment consists of a management unit which interfaces with flight management systems (FMSs) and a router which enables the aircraft to receive flight plans and weather information from the ground. 
The management unit enables the airline to update the FMS, and the crew can use this information to evaluate alternative flight plans while in flight.
\item {\textit{Ground  equipment:}} The ground equipment consists of networks of radio transceivers managed by a computer that handles the ACARS messages. 
\item {\textit{DSP:}}  The communication between the aircraft and the ground is linked via a data link managed by a datalink service provider (DSP); SITA and ARINC are the two primary providers. 
\end{enumerate}

\section{\label{sec:AdversariesThreat}Adversaries and Threat Analysis}
\subsection{\label{sec:threatactor}Threat Actors}
\begin{figure}[ht!]
    \centering
    \includegraphics[width=0.75\linewidth]{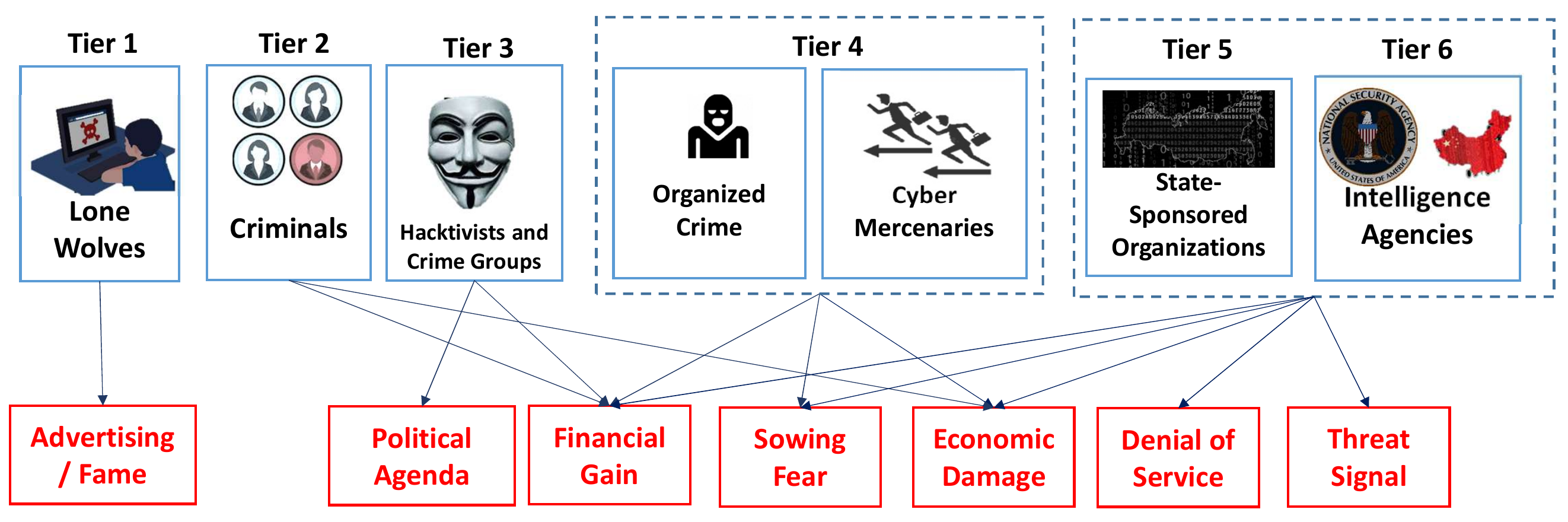}
    \caption{Primary adversaries and their goals.}
    \label{fig:ADV2}
\end{figure}

There is a wide range of possible adversaries; each type of adversary has a different motivation, purpose, and means at their disposal to achieve their goals. 
In this section, we review the different types of adversaries and their motives.
We consider an adversary as any individual or group who is trying to attack an airborne system in order to violate the security and proper functioning of the aircraft, 
We divided the adversaries into six tiers of based on their capabilities, potential for harm, and available resources.
Adversaries strive to achieve diverse goals, which are illustrated in Figure~\ref{fig:ADV2}. 
\begin{enumerate}[leftmargin=*]%[wide, labelwidth=!, labelindent=0pt]
\item{\textit{Tier 1:}} \textbf{`Lone Wolves'} are individuals who possess limited capabilities, knowledge, and techniques that are available on the internet. 
They are motivated by the possibility of gaining publicity, fame, or financial reward. 

\item{\textit{Tier 2:}} \textbf{Criminals} are individuals or teams of people who use technology to commit malicious activities on digital systems or networks in order to steal sensitive corporate data or personal information for profit.

\item {\textit{Tier 3:}} \textbf{Hacktivists} and \textbf{unorganized crime groups} are more advanced adversaries greater resources and knowledge. They are often motivated by political agendas, and their goal is usually to cause destruction and chaos.

\item {\textit{Tier 4:}} \textbf{Organized crime groups} and \textbf{cyber mercenaries} make up the fourth tier. Organized crime groups are usually driven by economic motives. 
These organizations can employ experts in the required fields and therefore can carry out attacks of a reasonable level of complexity. 
Their attacks are usually targeted at extortion and often include the threat of data disclosure or encryption using ransomware. 
Cyber mercenaries, employed by hostile stakeholders, may try to steal information or  damage property.
This can be achieved by stealing commercial or technological information or by compromising a company's reputation. 

\item {\textit{Tier 5:}} \textbf{State-sponsored organizations} form this tier of adversaries who represent the first level of so-called advanced persistent threats (APT). Such organizations are military organizations or technology groups funded by governments that aim to build strategic capabilities and increase accessibility to a wide range of targets. 
These organizations have unlimited resources and can devote significant time and high-quality human resources to gain powerful capabilities.
The targets of these organizations are carefully chosen, and their goals include delivering threat signals, causing operational and financial damage, causing denial of service or sowing fear.

\item {\textit{Tier 6:}} \textbf{Intelligence agencies} comprise the most advanced of the threat actor group. 
These adversaries have the offensive capabilities of a country's cyber military.
Like tier 5 threat actors, they have the most advanced capabilities and unlimited resources. 
Their goals range from gathering intelligence and business information to causing actual damage.
\end{enumerate}

\vspace{-8pt}
\subsection{\label{sec:Adversaries}Adversary Capabilities}
\begin{figure}[ht!]
    \centering
    \includegraphics[width=1\linewidth]{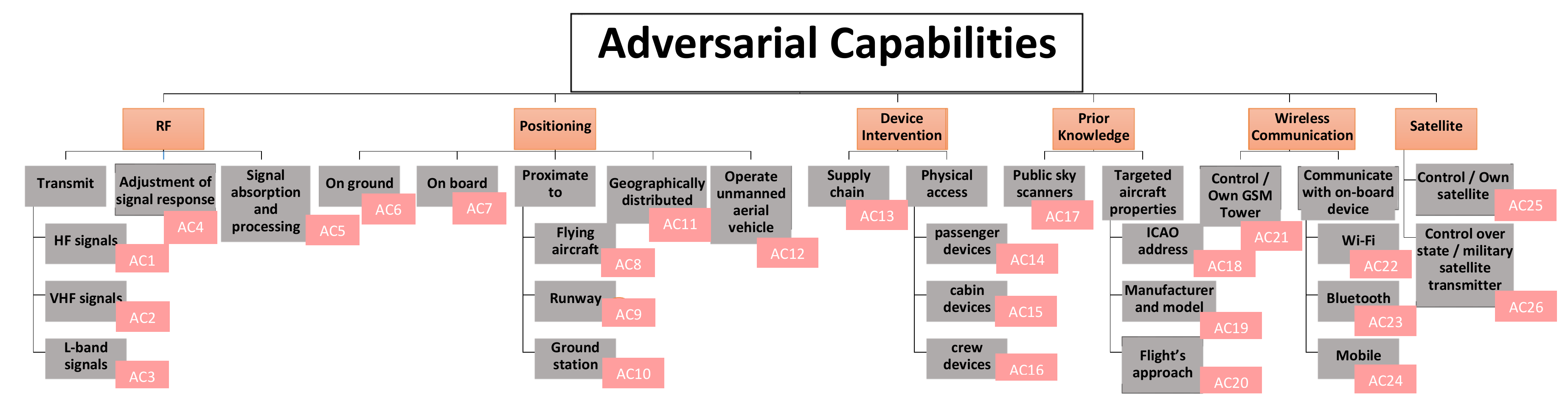}
     \caption{Adversarial capabilities.}
    \label{fig:CAPABILITIES}
\end{figure}
To better understand the capabilities required to implement the various attacks, we analyze the different adversarial capabilities (denoted as AC) adversaries need to execute various attacks. After examining the existing attacks on the multiple systems, we then classified the adversaries' capabilities based on the type of attack the capabilities facilitate.
Figure~\ref{fig:CAPABILITIES} illustrates the division of the capabilities into groups.
\begin{enumerate}[leftmargin=*]
\item\textbf{RF - Signaling capabilities }
\newline Radio frequency (RF) signaling refers to the capability of generating an electromagnetic signal, can be used as a type of communication. Radio waves are a form of electromagnetic radiation with identified radio frequencies that range from 3 kHz to 300 GHz.
RF communication is one of the primary means by which an aircraft communicates with its surroundings and ground stations. Therefore, the attacker's ability to transmit signals over specific ranges allows him/her to communicate and influence different systems.
    \begin{itemize}[wide, labelwidth=!, labelindent=0pt]
        \item \textit{AC1 - Transmit HF signals:} An adversary can operate in high frequency band (3-30 MHz), which is used by international shortwave radio stations in aviation communication. The HF system on an aircraft enables two-way voice communication with ground stations or other aircraft and provides digitally coded signals for such communication. 
        \item \textit{AC2 - Transmit VHF signals:} An adversary can operate in very high frequency band (30-300 MHz). Different systems use the frequencies in this range; For example, in aviation, the range of 108–118 MHz is used by the VOR, and the ILS localizer as air navigation beacons, while 118–137 MHz is used as airband for air traffic control, and 121.5 MHz serves as an emergency frequency.
        \item \textit{AC3 - Transmit L-band signals:} An adversary can operate at the L-band (1-2 GHz), which is the top end of the ultra high frequency (UHF) band. The L-band is used by various aircraft systems, e.g., the ADS-B, TCAS, and DME.
        \item \textit{AC4 - Adjust signal response:} The ability to sync the response to interrogations or adjust transmission rate. This capability is required to influence systems that depend on the signal's time of arrival to determine the transmitter's location.
        \item \textit{AC5 - Signal absorption and processing:} The adversary's ability to pick up, listen to, and decode signals using appropriate hardware (receivers), processors, and parsers (e.g., OpenSky data tools\footnote{https://opensky-network.org/data/data-tools} for processing ADS-B traffic).
    \end{itemize}
    
\item\textbf{Positioning capabilities}
\newline The adversary's position in relation to the attacked aircraft is significant in terms of the attacker's ability to carry out the different attacks. 
This importance derives from the systems' mode of operation, the degree of absorption, and the distance at which the threat actor operates.
Moreover, an adversary's location directly impacts the attack duration and accuracy, as the aircraft is not a static target.
    \begin{itemize}[wide, labelwidth=!, labelindent=0pt]
        \item \textit{AC6 - On ground:} An adversary located on the ground is limited in his/her ability to perform a prolonged attack on an object moving at high speed and high altitude. Ground attacks have the advantage of affecting a defined area.
        \item \textit{ AC7 - On board:} An adversary located on board has a high degree of destructive potential and long-term impact on the aircraft given his/her physical access to a variety of components. However, attackers are limited in terms of the means and tools that they can bring on board to perform the attack, and the degree of difficulty in concealing them is not trivial.
        \item \textit{AC8 - Proximate to flying aircraft:}  Proximity to the aircraft during flight enables the execution of attacks that depend on close distance to the aircraft (e.g., TCAS spoofing and jamming).
        \item \textit{AC9 - Proximate to the runway:} Proximity to the runway is required to cause interference during the aircraft's landing and take-off (e.g.,  Proximity to the runway is required for the aim of abusing the ILS  whose operation activates upon the landing phase).
        The risk inherent in runway proximity is the high chance of being caught due to the presence of security personnel, identification systems, etc. 
        \item \textit{AC10 - Proximate to the ground station:} Proximity to the ground station can enable the disruption of frequencies near the station and facilitate impersonation attacks by taking advantage of the physical proximity to a legitimate station.
        \item\textit{AC11 - Geographically distributed:} A decentralized adversary can perform a synchronized attack in several places at a coordinated time or simultaneously in order to affect a target or a comprehensive area partition. For example, using the multilateration concept, an adversaries can execute spoofing of GPS signals while they are distributed between different areas.
        \item \textit{AC12 - Operate unmanned aerial vehicle:} Similar to the AC8 capability, operating an unmanned aerial vehicle enables proximity to an aircraft during flight.
        This route to proximity has several advantages, e.g.,  it reduces the adversaries exposure.
    \end{itemize}
    
\item\textbf{Device intervention capabilities }
\noindent\newline Device intervention refers to the ability to affect components and systems on the aircraft through direct or indirect access for the purpose of disrupting or injecting malicious payloads or backdoors.
    \begin{itemize}[wide, labelwidth=!, labelindent=0pt]
        \item \textit{AC13 - Supply chain:} The ability to influence the supply chain of various technologies in aircraft in order to introduce infected components. This attack has vast potential for damage, and its implementation requires extensive knowledge of deployment and procurement processes.
        \item \textit{AC14 - Physical access to passenger's device:} Accessibility to passenger's devices can be used in order to harm components in the PIESD and PODD (e.g., the IFEC system).
        \item \textit{AC15 - Physical access to cabin devices:} Accessibility to devices within the cabin can be used in order to harm components in the PIESD and AISD (e.g., the FDR).
        \item \textit{AC16 - Physical access to crew devices: }Accessibility to the crews' devices can be used in order to harm components in the AISD and ACD (e.g., the EFB).
    \end{itemize}
\item\textbf{Prior knowledge capabilities }
\newline Prior knowledge is a prerequisite for targeted attacks; gaining information about an aircraft and its trajectory can highly influence how successful and effective an attack can be.
    \begin{itemize}[wide, labelwidth=!, labelindent=0pt]
        \item \textit{AC17 - Public sky scanners: } Access to public databases and scanners, such as OpenSky Network and Flightradar24, enables access to real-time and historical information regarding flight routes and aircraft themselves.
        \item\textit{AC18 - Obtain ICAO address:} Aircraft are assigned a unique ICAO 24-bit address upon national registration, which becomes a part of the aircraft's certificate of registration. Since this address normally never changes, obtaining this information allows an adversary to target a specific aircraft and track its route.
        \item \textit{AC19 - Obtain manufacturer and model information:} Obtaining information concerning the aircraft manufacturer and model can provide the adversary with details regarding the versions of aircraft components and the technologies deployed. An adversary can use such information to detect attack surfaces and design the most suitable vector to achieve its goals.
        \item \textit{AC20 - Flight's approach route}: Obtaining the flight's approach route and the flight plan of a targeted aircraft can help an adversary determine the best location to carry out an attack. 
        This type of information is necessary for the interference with the landing phase (e.g., ILS spoofing, and GBAS).
    \end{itemize}
\item\textbf{Wireless communication capabilities }
\newline Aircraft systems communicate remotely with satellites and ground stations. To access these systems, an adversary must possess the ability to communicate via wireless communication.

    \begin{itemize}[wide, labelwidth=!, labelindent=0pt]
        \item \textit{AC21 - Control/own GSM tower:} Owning a GSM tower (e.g., cell tower) is required for cellular communication with third-party service providers and onboard systems (e.g., IFEC). 
        \item \textit{AC22 - Communicate with an onboard device using Wi-Fi:} Gaining Wi-Fi capabilities can be used for communicating of onboard devices and even affect them; e.g., Wi-Fi vulnerabilities can be used to spread between devices connected to the same Wi-Fi network. For instance, the airport Wi-Fi network can be used as an attack surface to infect pilots' and crew's devices (e.g., the EFB terminal).
        \item \textit{AC23 - Communicate with an onboard device using Bluetooth:} Similar to AC22, an adversary can use Bluetooth exploits to infect nearby components (e.g., passengers' cellular devices).
        \item \textit{AC24 - Communicate with an onboard device using mobile communications:} Using cellular communications, an adversary can maintain continuous communication with components on the aircraft using mobile-satellite services (e.g., Viasat and Intelsat).
    \end{itemize}
\item\textbf{Satellite capabilities}
\newline Capabilities dealing with satellites are often powerful capabilities associated with state or military entities. The use of satellite capabilities can provide a wide range of attack surfaces and facilitate complex attacks that require high transmission intensities.
    \begin{itemize}[wide, labelwidth=!, labelindent=0pt]
        \item \textit{AC25 - Control/own satellite:} An adversary controlling a satellite can obtain accurate information from a wide geographical surface. In addition, such control provides the ability to transmit powerful signals in order to disrupt communication channels and give inaccurate information to different stakeholders (e.g., aircraft, and ground stations).
        \item \textit{AC26 - Control state/military satellite transmitter:} A powerful GPS transmitter can be used for GPS spoofing and jamming attacks. Moreover, the military SATCOM transmitter provides secure and reliable connectivity, and can reach a wide range using strong amplifiers.
    \end{itemize}
\end{enumerate}

Table~\ref{tab:AdversaryCapabilities} provides a comparison of the different tiers of threat actors based on their feasibility of acquiring different capabilities to implement an attack. 
The index is the degree of likelihood of possessing a capability which ranges from certian feasibility to improbable feasibility.

\begin{center}

\begin{table}[ht!]
%\hspace*{-1.3cm}
\centering
\tiny
\begin{tabular}{
|m{0.06\textwidth}|NNNNN|NNNNNNN|NNNN|NNNN|NNNN|NN|}
\Xhline{2pt}

\multirow{3}{*}{\textbf{Adversary}} &
\multicolumn{25}{c}{\textbf{\underline{Adversary Capabilities}}} &
\\ 
    &
    \multicolumn{5}{c}{\textbf{\underline{RF}}} &
    \multicolumn{7}{c}{\textbf{\underline{Positioning}}} &
    \multicolumn{4}{c}{\textbf{\underline{ Intervention}}} &
    \multicolumn{4}{c}{\textbf{\underline{ Knowledge}}} &
    \multicolumn{4}{c}{\textbf{\underline{ Comm}}} &
    \multicolumn{2}{c|}{\textbf{\underline{Satellite}}} \\
    & \rotatebox{90}{HF signals}
    & \rotatebox{90}{VHF signals}
    & \rotatebox{90}{L-band signals}
    & \rotatebox{90}{Adjustment}
    & \rotatebox{90}{Absorption / processing}
    & \rotatebox{90}{Onground}
    & \rotatebox{90}{Onboard}
    & \rotatebox{90}{Proximate-Flying Aircraft}
    & \rotatebox{90}{Proximate-Runway}
    & \rotatebox{90}{Proximate-GS}
    & \rotatebox{90}{Distributed}
    & \rotatebox{90}{Operate UAV}
    & \rotatebox{90}{Supply chain}
    & \rotatebox{90}{Access to passenger device}
    & \rotatebox{90}{Access to cabin device}
    & \rotatebox{90}{Access to crew device}
    & \rotatebox{90}{Sky scanner}
    & \rotatebox{90}{ICAO address}
    & \rotatebox{90}{Manufacturer data}
    & \rotatebox{90}{Approach route}
    & \rotatebox{90}{control over GSM Tower}
    & \rotatebox{90}{Wi-Fi}
    & \rotatebox{90}{Bluetooth}
    & \rotatebox{90}{Mobile}
    & \rotatebox{90}{Control over satellite}
    & \rotatebox{90}{Own military transmitter} \\
\Xhline{2pt}
\tiny{Tier 1}
    & $\bullet$ %SATCOM 
    & $\bullet$ %CPDLC   
    & $\bullet$ %ACARS  
    & $\bullet$ %DME 
    & $\bullet$ %VOR 
    & $\bullet$%NDB
    & $\halfcirc$%ILS
    & $\circ$%DNNSS
    & $\circ$%ABAS
    & $\circ$%SBAS
    & $\circ$%GNAS
    & $\circ$%PSR
    & $\circ$%SSR
    & $\circ$%MLAT
    & $\circ$%ADS-B
    & $\circ$%TCAS
    & $\bullet$%ACAS X
    & $\bullet$%Engine Alerting
    & $\halfcirc$%FIS-B
    & $\bullet$%TIS-B
    & $\circ$%EFB
    & $\bullet$%TWLU 
    & $\bullet$%CWLU
    & $\bullet$%EGM
    & $\circ$%NIM
    & $\circ$%CSS
    \\\hline
    
\tiny{Tier 2}
    & $\bullet$ %SATCOM 
    & $\bullet$ %CPDLC   
    & $\bullet$ %ACARS  
    & $\bullet$ %DME 
    & $\bullet$ %VOR 
    & $\bullet$%NDB
    & $\halfcirc$%ILS
    & $\circ$%DNNSS
    & $\circ$%ABAS
    & $\circ$%SBAS
    & $\bullet$%GNAS
    & $\circ$%PSR
    & $\circ$%SSR
    & $\circ$%MLAT
    & $\circ$%ADS-B
    & $\circ$%TCAS
    & $\bullet$%ACAS X
    & $\bullet$%Engine Alerting
    & $\halfcirc$%FIS-B
    & $\bullet$%TIS-B
    & $\circ$%EFB
    & $\bullet$%TWLU 
    & $\bullet$%CWLU
    & $\bullet$%EGM
    & $\circ$%NIM
    & $\circ$%CSS
    \\\hline
\tiny{Tier 3}
    & $\bullet$ %SATCOM 
    & $\bullet$ %CPDLC   
    & $\bullet$ %ACARS  
    & $\bullet$ %DME 
    & $\bullet$ %VOR 
    & $\bullet$%NDB
    & $\bullet$%ILS
    & $\bullet$%DNNSS
    & $\circ$%ABAS
    & $\circ$%SBAS
    & $\bullet$%GNAS
    & $\bullet$%PSR
    & $\circ$%SSR
    & $\bullet$%MLAT
    & $\circ$%ADS-B
    & $\circ$%TCAS
    & $\bullet$%ACAS X
    & $\bullet$%Engine Alerting
    & $\halfcirc$%FIS-B
    & $\bullet$%TIS-B
    & $\circ$%EFB
    & $\bullet$%TWLU 
    & $\bullet$%CWLU
    & $\bullet$%EGM
    & $\circ$%NIM
    & $\circ$%CSS
    \\\hline
\tiny{Tier 4}
    & $\bullet$ %SATCOM 
    & $\bullet$ %CPDLC   
    & $\bullet$ %ACARS  
    & $\bullet$ %DME 
    & $\bullet$ %VOR 
    & $\bullet$%NDB
    & $\bullet$%ILS
    & $\bullet$%DNNSS
    & $\circ$%ABAS
    & $\circ$%SBAS
    & $\bullet$%GNAS
    & $\bullet$%PSR
    & $\halfcirc$%SSR
    & $\bullet$%MLAT
    & $\halfcirc$%ADS-B
    & $\bullet$%TCAS
    & $\bullet$%ACAS X
    & $\bullet$%Engine Alerting
    & $\bullet$%FIS-B
    & $\bullet$%TIS-B
    & $\bullet$%EFB
    & $\bullet$%TWLU 
    & $\bullet$%CWLU
    & $\bullet$%EGM
    & $\circ$%NIM
    & $\circ$%CSS
    \\\hline
\tiny{Tier 5}
    & $\bullet$ %SATCOM 
    & $\bullet$ %CPDLC   
    & $\bullet$ %ACARS  
    & $\bullet$ %DME 
    & $\bullet$ %VOR 
    & $\bullet$%NDB
    & $\bullet$%ILS
    & $\bullet$%DNNSS
    & $\bullet$%ABAS
    & $\bullet$%SBAS
    & $\bullet$%GNAS
    & $\bullet$%PSR
    & $\halfcirc$%SSR
    & $\bullet$%MLAT
    & $\bullet$%ADS-B
    & $\bullet$%TCAS
    & $\bullet$%ACAS X
    & $\bullet$%Engine Alerting
    & $\bullet$%FIS-B
    & $\bullet$%TIS-B
    & $\bullet$%EFB
    & $\bullet$%TWLU 
    & $\bullet$%CWLU
    & $\bullet$%EGM
    & $\bullet$%NIM
    & $\bullet$%CSS
    \\\hline
\tiny{Tier 6}
    & $\bullet$ %SATCOM 
    & $\bullet$ %CPDLC   
    & $\bullet$ %ACARS  
    & $\bullet$ %DME 
    & $\bullet$ %VOR 
    & $\bullet$%NDB
    & $\bullet$%ILS
    & $\bullet$%DNNSS
    & $\bullet$%ABAS
    & $\bullet$%SBAS
    & $\bullet$%GNAS
    & $\bullet$%PSR
    & $\bullet$%SSR
    & $\bullet$%MLAT
    & $\bullet$%ADS-B
    & $\bullet$%TCAS
    & $\bullet$%ACAS X
    & $\bullet$%Engine Alerting
    & $\bullet$%FIS-B
    & $\bullet$%TIS-B
    & $\bullet$%EFB
    & $\bullet$%TWLU 
    & $\bullet$%CWLU
    & $\bullet$%EGM
    & $\bullet$%NIM
    & $\bullet$%CSS
    \\\Xhline{2pt}
\end{tabular}

\caption{Comparison between the ability of the different threat actors to achieve the various capabilities.
\centering
CERTIAN FEASIBILITY ( $\bullet$ ), REASONABLE FEASIBILITY ($\halfcirc$), IMPROBABLE FEASIBILITY  ($\circ$)}
\label{tab:AdversaryCapabilities}
\end{table}
\end{center}

%\clearpage
\section{\label{sec:cybersecurity} Concrete Avionics Attacks}
In this section, we review existing cyber attacks on the various avionics systems, both in academia and industry.

\begin{enumerate}[wide, labelwidth=!, labelindent=0pt]
\item \textbf{ADS-B system cyber attacks:}
The ADS-B system lacks basic security mechanisms such as authentication, message integrity, and encryption. 
Because it is used to provide information in real time, these security gaps make the application of the protocol in crowded skies risky, exposing the aircraft to different types of attacks.
Costin \textit{et al.}~\cite{costin2012ghost} showed that jamming, denial-of-service (DoS), eavesdropping, spoofing, and impersonation attacks are both easy and practically
feasible for a moderately sophisticated attacker to apply on the ADS-B system. 
Both \emph{FIS-B} and \emph{TIS-B} may be susceptible to similar attacks, since they are all transmitted
over an unauthenticated link and carry the same data format.
The ease with which these attacks can be executed using COTS transponders has also been described in~\cite{eskilsson2020demonstrating}.
A description of attack trees that describe the steps that need to be taken to implement the various attacks is provided in~\cite{viveros2016analysis}, and attack scenarios are listed in~\cite{mirzaei2019security}.
\item \textbf{SSR system cyber attacks:}
The SSR system is prone to spoofing, jamming, and over-interrogation and radar loss.
A major vulnerability of the SSR system is the ability to compromise its availability, whether intentionally or unintentionally, by overloading interrogations ~\cite{strohmeier2020securing}.
As high interrogation rates can cause transponders to restrict their sensitivity to interrogations and overheat, and result in radar loss of the target from ATC displays.
Moreover, Osechas \textit{et al.}~\cite{osechas2017addressing} and Mostafa \textit{et al.}~\cite{mostafa2016vulnerability} both showed that SSR systems are prone to jamming attacks using a high-power transmitter at frequencies of 1030 MHz for interrogation and 1090 MHz for replies, and introduced and introduced different scenarios of spoofing attacks, e.g. injecting ghost aircraft into display screens. 
\item \textbf{MLAT system cyber attacks:}
As a derivative of the MLAT mode of operation, the MLAT system is not affected by tampering attacks, since the only necessary information is the signal's time of arrival.
In contrast, GPS spoofing technique have already been discussed in the literature, where they were shown to be a potential attack vector of the MLAT system. Moser \textit{et al.}~\cite{moser2016investigation} demonstrated how a distributed and multi-instrument attacker could disrupt the system and spoof aircraft positions.
\item \textbf{TCAS/ACAS X cyber attacks:}
The S system is designed to reduce the incidence of mid-air collisions with other aircraft, therefore an attack that disrupts the system's operation or gives the attacker control of the system is a risk to human life.
Hannah \textit{et al.}~\cite{hannah2021cyber} provided a description of the general landscape of the S threat actors, their capabilities, and potential attacks. 
Furthermore, both the SESAR and NextGen projects, which are the leading modernization efforts in the aviation industry, plan to implement new operational concepts which will reduce the space between aircraft. In addition, the FAA has funded research and development on ACAS X, which will likely replace the S system. 
Research~\cite{smith2020understanding} has shown how ACAS X can be triggered to erode the safety of the aircraft with the use of expensive equipment and distance constraints.
These attacks are difficult to implement, however they are particularly dangerous because once they have been successfully implemented, even an experienced pilot will have difficulty identifying the attack~\cite{smith2020view}.
\item \textbf{Engine alerting system cyber attacks:}
The EICAS is used to display engine parameters and raises alerts regarding configuration or faults.
Therefore, exploitation of the EICAS/ECAM system can be used to manipulate the crew's behavior and affect the engine. 
Chris Roberts, a security researcher, claimed that he was able to spoof EICAS messages using unsupervised access to the FMS when he was on a flight; his claim was published by Kaspersky Labs~\cite{EICASKaspersky}.
\item \textbf{GNSS cyber attacks:}
As aviation operations increasingly rely on the GNSS to improve navigation performance and support air traffic control surveillance functions, vulnerabilities in it have the potential to cause widespread damage.
Industry and academia have shown how jamming, intentional disruptions, and spoofing can influence the GNSS by utilizing the frequencies at which it operates (L1/L2), thus affecting many applications that use satellite information for the purpose of obtaining precise directions or location.
Truffer\textit{et al.}~\cite{truffer2017jamming} illustrated how GNSS jamming can affect the position displayed on the FMS. In addition, Tanil \textit{et al.}~\cite{tanil2016ins} discussed the potentially catastrophic impact of GNSS spoofing at remote locations, where traditional ILS services are unavailable and the landing approach depends on Ground-Based Augmentation Systems (GBASs).
\item \textbf{Ground-based navigation system cyber attacks:}
\begin{itemize}[leftmargin=*]
\item The ILS is a radio navigation system that provides short-range guidance to the aircraft, therefore exploitation of the system requires the attacker to be located near the aircraft, a requirement that puts the attacker at risk of detection.
ILS spoofing was introduced at DEFCON~\cite{ILSandSSpooging} where it was shown that a successful attack requires the placement of a powerful antenna in very close proximity to the airport.
There are few examples in the literature of possible wireless attacks on the ILS, two of which were described in~\cite{sathaye2019wireless}: an overshadow attack
and a single tone attack. The overshadow attack requires the attacker to overpower legitimate ILS signals, which causes the receiver to process the attacker's signal. In a single tone attack, an attacker transmits a single frequency tone signal at a lower strength than the legitimate ILS signal thus interfering with the original signal.
A successful ILS attack can disrupt the aircraft's ability to land safely and can therefore result in property damage, injury, and even death.
\item The VOR system is prone to jamming and spoofing attacks~\cite{choudhary2022aviation}, but with the development of precision approach systems, the use of non-precision approach systems such as VOR and NDB has significantly decreased today; therefore the threat posed by attacking this system does not have the potential for much damage.
\end{itemize}
\item \textbf{IFEC system cyber attacks:}
The IFEC system is more troublesome in security contexts, as the system is directly accessible to the passenger and therefore prone to breaches.
The IFEC system contains passengers' private information, and an adversary exploiting the system can gain control of the information passengers presented on their in-flight screen. 
Moreover, the system is connected to the Wi-Fi network and network controller that connects the PIESD to the ACD network. 
Exploiting the IFEC could allow an attacker to pass from the PIESD to more sensitive networks.
In~\cite{IFECIOACTIVE} and~\cite{IFECsystem}, the researchers showed how the IFEC system's vulnerabilities can be used to enable the attacker to pass between networks, access credit card details, and control cabin lighting and smart screens.
\item \textbf{SATCOM cyber attacks:}
The aircraft's SATCOM system datalink serves multiple systems of the aircraft's control and crew devices; the datalink is used for voice and data services, allowing an aircraft to communicate via satellite.
An exploit targeting the SATCOM infrastructure (protocol, devices, services) can be used to gain remote control of various systems.
The severity of SATCOM exploitation used to access passenger and crew devices was discussed in~\cite{santamarta2018last}, while Santamarta \textit{et al.}~\cite{santamarta2014satcom,santamarta2014wake} described how an adversary can abuse SATCOM terminals to find a backdoor and retrieve hardcoded credentials. 
The authors claim that an adversary exploiting SATCOM terminals has the potential to intercept, manipulate, and block communications, and in some cases, to remotely take control of the physical device.
\item \textbf{CPDLC system cyber attacks:}
The CPDLC system is unencrypted and therefore does not meet basic security and privacy requirements.
Gurtov \textit{et al.}~\cite{gurtov2018controller} analyzed the CPDLC system's technical features and properties and divided the possible threat actors into active and passive threat actors. The authors describe how CPDLC system exploits can be used for eavesdropping, jamming, flooding, injection, alteration, and masquerading attacks thus enabling an attacker to gain access and control messages, modify their content, and flood ground stations and aircraft with ghost messages. 
Di Macro \textit{et al.}~\cite{di2016security} presented a more sophisticated attack in which CPDLC systems can be attacked through a man-in-the-middle (MITM) attack with the use of open-source tools. In ~\cite{smailes2021you}, the author expanded on the MITM attack and explained how it can be used to take over an aircraft's communications and transmit CPDLC commands without alerting the legitimate controller. The feasibility of transmitting crafted CPDLC messages was discussed by Eskilsson in~\cite{eskilsson2020demonstrating}.
\item \textbf{ACARS cyber attacks:}
The ACARS was developed with no security measures~\cite{smith2016security}. 
In recent years, several attack scenarios have been demonstrated by both security researchers and hackers in industry, e.g., in 2012, security researcher Hugo Teso showed how malicious ACARS messages can be crafted by an adversary and used to control the flight management system and thereby also control the pilot's displays and control systems, using just a simple mobile phone~\cite{teso2013aircraft}. 
An introduction to cyber attacks on ACARS was also presented at DEFCON~\cite{IntroToAcars}; one of the main points raised was the threat posed by the existing physical links between the communication management unit, which is used to route ACARS traffic, and the various avionics systems, whereby a vulnerability in the ACARS has the ability to affect many other systems.
In~\cite{ImpactAssessment}, the European Union Aviation Safety Agency (EASA) addressed the ACARS cyber security threats, analyzing two scenarios: weight and balance update events and flight plan update events.
The weight and balance update attack deals with an onground attacker who sends crafted ACARS updated to the aircraft, which can result in uncontrollable behavior of the aircraft.
The flight plan update attack involves an attacker who transmits falsified flight plan data to a targeted aircraft; in this case, a successful custom attack requires prior knowledge regarding the aircraft's route.
This type of attack can result in deviations from the desired route.
\item \textbf{Loadable cyber attacks:}
Aircraft systems can be modified using loadable software; this allows their configuration to be updated without physical intervention. Modifications and replacements can be made via remote wireless services or pluggable devices.
Security researchers demonstrated they can interfere with the update process by impersonating a legitimate operator or achieving unauthorized physical access. With the ability to update loadable software, they showed how the navigation database~\cite{Loadbles2} on the flight management system can be manipulated, as well as how an attacker can take control of onboard systems using an AMI wireless data loader~\cite{Loadbles1}.
\item \textbf{EFB cyber attacks:}
The EFB connects to multiple systems and has the ability to access those systems and multiple applications, including: the flight management system, passenger information list, performance applications, technical logs, weight and balance applications, and flight planning application.
The portable nature of the EFB and its ability to connect to public networks put the EFB tablet at high risk. 
For example, exposure to public networks may allow the system to be attacked through Wi-Fi vulnerabilities~\cite{IFECsystem}. 
Despite the importance and vulnerability of the system, surveys show that most airlines do not have a cyber security plan in place for the tablet-based EFB used by their pilots~\cite{EFBReport}.
An example of attack vectors for connecting the FMS, retrieving sensitive data, accessing flight planning and navigation applications, and even modifying weight and balance calculations are presented in~\cite{EFBPPT}.
\end{enumerate}

Summarizing the material presented in this section, Table~\ref{tab:CIATriangle} contains a list of the attacks known to academia and industry with a short description of the sub-technique that was carried out as part of the attack.
We used the STRIDE~\cite{shostack2006uncover} threat model to group the attacks into different categories. This model considers six threat categories (spoofing, tampering, repudiation, information disclosure, denial of service, and elevation of privileges).
In addition, the table includes references for academic studies and industry implementations; These are the \emph{procedures} in which the various cyber attacks are described and implemented, their required capabilities (listed in Section~\ref{sec:Adversaries}), exploited vulnerabilities, and the identified threat posed by the attack according to the STRIDE threat model.

\begin{landscape}
\renewcommand{\arraystretch}{1.3}
{\scriptsize
\begin{longtable}[t]{|p{0.08\textwidth}|p{0.17\textwidth}|p{0.15\textwidth}|p{0.29\textwidth}|p{0.15\textwidth}|p{0.1\textwidth}|p{0.15\textwidth}|p{0.15\textwidth}|} 
\caption{Risk assessment of aviation systems - known attacks, impacts, vulnerabilities, required capabilities, target assets and identified threats.}
\label{tab:CIATriangle}\\
\Xhline{2pt}
\textbf{Asset}& \textbf{Technique}& \textbf{Procedures}& \textbf{Sub-Technique}  & \textbf{Impact}   & \textbf{Capabilities}& \textbf{Vulnerabilities}& \textbf{STRIDE}
\\ \Xhline{2pt}
\endhead
\multirow{5}{*}{ADS-B}& Aircraft reconnaissance& \multirow{5}{*}{\cite{viveros2016analysis,costin2012ghost,mirzaei2019security,eskilsson2020demonstrating}}& Receive, parse, and collect~messages with a radio receiver& Aircraft tracking & AC5 , AC6, AC17& No Confidentiality& Information disclosure   
\\ \cline{2-2}\cline{4-8}
  & Aircraft flood denial&  & Create destructive signal interference (1090 MHz)   & Loss of view& AC3 , (AC6 | AC7)& No Availability   & Denial of service  
\\ \cline{2-2}\cline{4-8}
  & Ghost aircraft injection   &  & Transmit signal that conforms to a protocol and mirrors legitimate traffic  & Manipulation of view& AC3 , (AC6 \textbar{} AC7)& No Authentication& Spoofing
\\ \cline{2-2}\cline{4-8}
  & Virtual trajectory modification  &  & Bit-flipping, overshadowing~in order to modify ADS-B messages  & Manipulation of control & AC3 , AC5, (AC6 \textbar{} AC7) & No Integrity& Tampering
\\ \cline{2-2}\cline{4-8}
  & Aircraft disappearance&  & Transmit destructive or constructive interference   & Loss of view& AC3 , AC5, (AC6 \textbar{} AC7) & \multirow{2}{*}{No Availability}   & \multirow{2}{*}{DoS}
\\ \Xhline{2pt}
\multirow{2}{*}{MLAT}& Missynchronization & \multirow{2}{*}{\cite{hering2003safety,moser2016investigation}}& Block GPS signals to interfere with MLAT ground receiver synchronization  & DoS – Pilot & (AC2 \textbar{} AC3), AC11&No Authentication& Spoofing  
\\ \cline{2-2}\cline{4-8}
  & Spoof locations&  & Use multiple devices to spoof GPS signals& Manipulation of view& (AC2 \textbar{} AC3), AC11& No Authentication& Spoofing
\\ \Xhline{2pt}
\multirow{4}{*}{SSR}& Overloading-interrogations & \multirow{4}{*}{\cite{strohmeier2020securing,osechas2017addressing,mostafa2016vulnerability}}& Flood interrogations at 1030 MHz to exceed the acceptable standard& \multirow{2}{*}{Loss of view} & AC3, (AC6 \textbar{} AC7) & No Availability   & DoS
\\ \cline{2-2}\cline{4-4}\cline{6-8}
  & Block SSR signals&  & Delete SSR transmissions  & & AC3, (AC6 \textbar{} AC7), AC18 & No Authentication& Spoofing
\\ \cline{2-2}\cline{4-8}
  & Ghost aircraft &  & Spoof fake SSR transmissions & \multirow{2}{*}{Manipulation~of view}& AC3, (AC6 \textbar{} AC7) & \multirow{2}{*}{No Integrity}& \multirow{2}{*}{Tampering}
\\ \cline{2-2}\cline{4-4}\cline{6-6}
  & Tampering with SSR transmission &  & Use an SDR that an attacker can alter, block, and inject Mode A, Mode B, Mode C, and Mode S messages  & & AC3, (AC6 \textbar{} AC7) &&  
\\ \Xhline{2pt}
\multirow{2}{*}{ILS}& Overshadow& \multirow{2}{*}{\cite{sathaye2019wireless,ILSandSSpooging}}& Overpower legitimate signals~using specially crafted ILS signals& \multirow{2}{*}{Loss of safety}& \multirow{2}{*}{\makecell[l]{AC2, AC6, AC9 \\AC19, AC20}}  & \multirow{2}{*}{No Availability}   & \multirow{2}{*}{Spoofing}
\\ \cline{2-2}\cline{4-4}
  & Single tone&  & Cause deflections in the course deviation indicator needle& &   &&  
\\ \Xhline{2pt}
VOR& Aircraft reconnaissance& \cite{mostafa2016vulnerability,sathaye2019wireless,choudhary2022aviation} & Receive, parse, and collect~messages with a radio receiver& Aircraft tracking & AC2 AC5 & No Confidentiality& Information disclosure   
\\ \Xhline{2pt}
\multirow{3}{*}{\makecell[l]{GNSS\\SBAS\\ABAS\\GBAS}} & Disrupt landing approach   & \multirow{3}{*}{\cite{tanil2016ins,ochieng2003gps,pollack2018aviation,truffer2017jamming}}& Spoof GNSS and GNSS augmentation(SBAS, ABAS, GBAS) system using GPS signals& Loss of safety& \multirow{3}{*}{\makecell[l]{AC3, AC18\\AC25, AC26}}& \multirow{2}{*}{No Availability}   & \multirow{2}{*}{DoS}
\\ \cline{2-2}\cline{4-5}
  & Deceive a GPS receiver&  & Transmit a slightly more powerful signal~than that received from the GPS satellites& \multirow{2}{*}{Manipulation of view}&   &&  
\\ \cline{2-2}\cline{4-4}\cline{7-8}
  & Display unreliable position information&  & Transmit interference signals to affect displayed FMS position data   & &   & No Integrity& Tampering
\\ \Xhline{2pt}
DME& Aircraft reconnaissance& \cite{hagmuller2004speech}   & Receive, parse, and collect~messages with a radio receiver& Aircraft tracking & (AC2 \textbar{} AC3), AC5, AC10 & No Confidentiality& Information disclosure   
\\ \Xhline{2pt}
\multirow{5}{*}{ACARS}& Aircraft reconnaissance& \multirow{5}{*}{\cite{smailes2021you,di2016security,eskilsson2020demonstrating,IntroToAcars,smith2016security,teso2013aircraft}}  & Receive, parse, and collect ATC, AOC,AAC, OOOI messages with a radio receiver& Aircraft tracking & AC5, AC6& No Confidentiality& Information disclosure   
\\ \cline{2-2}\cline{4-8}
  & Bogus flight plan update   &  & \multirow{4}{*}{\makecell[l]{Issue and transmit malicious message\\(ATC, AOC,AAC) that conforms to a\\protocol and mirrors legitimate ACARS\\traffic}} & \multirow{2}{*}{Loss of control}& (AC1\textbar{}AC2), AC6, AC18, AC20   & \multirow{4}{*}{No Authentication} & \multirow{4}{*}{Spoofing}
\\ \cline{2-2}\cline{6-6}
  & Weight and balance manipulation  &  && & (AC1\textbar{}AC2), AC6,~AC18, AC19   &&  
\\ \cline{2-2}\cline{5-6}
  & Malicious requests for~passenger information&  && Theft of passengers' information& \multirow{2}{*}{\makecell[l]{(AC1\textbar{}AC2),\\AC6}}&&  
\\ \cline{2-2}\cline{5-5}
  & Distort weather information&  && Loss of view&   &&  
\\ \Xhline{2pt}
\multirow{7}{*}{CPDLC}& Read and collect~control messages& \multirow{7}{*}{\cite{costin2012ghost,mirzaei2019security,eskilsson2020demonstrating,smith2017analyzing}}& Receive, parse, and collect messages with a radio receiver & Theft of aircraft statistics  & AC5& No Confidentiality& Information disclosure   
\\ \cline{2-2}\cline{4-8}
  & Impersonate an ATC&  & Craft and inject messages claiming to be a CPDLC unit on the ATC end & Manipulation of control   & AC2, AC18& No Authentication& Spoofing
\\ \cline{2-2}\cline{4-8}
  & Disrupt communication between aircraft and an ATSU &  & Block legitimate messages and compromise~an ongoing CPDLC connection handover& \multirow{2}{*}{DoS - Pilot}  & \multirow{3}{*}{\makecell[l]{AC2, AC8,\\ AC18}} & No Integrity& Tampering
\\ \cline{2-2}\cline{4-4}\cline{7-8}
  & Selective message jamming  &  & Block session termination message & &   & \multirow{2}{*}{No Availability}   & \multirow{2}{*}{DoS}
\\ \cline{2-2}\cline{4-5}
  & Aircraft flood denial&  & Reduce CPDLC channel's~capacity by filling the channel with noise& Isolation from operating factors&   &&  
\\ \cline{2-2}\cline{4-8}
  & Aircraft ghost messaging   &  & Transmit unauthorized CPDLC messages& \multirow{2}{*}{\makecell[l]{Manipulation of\\control}}  & AC2, AC18& \multirow{2}{*}{No Authentication} & \multirow{2}{*}{Spoofing}
\\ \cline{2-2}\cline{4-4}\cline{6-6}
  & Issue incorrect commands   &  & Transmit unauthorized CPDLC messages& & AC2, AC18, AC19&&  
\\ \Xhline{2pt}
\multirow{3}{*}{SATCOM}   & Disrupt, intercept, or modify in-flight Wi-Fi  & \multirow{3}{*}{\cite{santamarta2014satcom,santamarta2014wake,santamarta2018last}}  & Block traffic from SATCOM~direct router to cause a denial of service   & DoS - Passengers  & \multirow{3}{*}{\makecell[l]{AC2,\\(AC25\textbar{}AC26)}}& \multirow{3}{*}{No Availability}   & \multirow{3}{*}{DoS}
\\ \cline{2-2}\cline{4-5}
  & Control crew and passenger devices   &  & Craft and inject information consumed by passengers' devices& DoS - Passengers and crew&   &&  
\\ \cline{2-2}\cline{4-5}
  & Compromise aircraft SATCOM to block traffic  &  & Transmit SATCOM communication~using destructive interference& Isolation from operating factors&   &&  
\\ \Xhline{2pt}
\multirow{3}{*}{Loadables}& Inject false data to the FMS   & \multirow{3}{*}{\cite{Loadbles1,Loadbles2}}  & Load configurations to the FMS& \multirow{3}{*}{\makecell[l]{Manipulation of \\control}}  & \multirow{3}{*}{\makecell[l]{AC13 \\\textbar{} (AC14, AC15\\, AC16)\\(AC22\textbar{} AC23\\\textbar{}AC24)}} & \multirow{3}{*}{No Integrity}& \multirow{3}{*}{Tampering}
\\ \cline{2-2}\cline{4-4}
  & Update the FMC with~corrupted navigation data&  & Load configurations to the FMCS and NDB & &   &&  
\\ \cline{2-2}\cline{4-4}
  & Remotely update malicious AMI&  & Load malicious malformed airline modifiable information   & &   &&  
\\ \Xhline{2pt}
\multirow{4}{*}{\makecell[l]{TCAS\\ACAS X}}  & Aircraft reconnaissance& \multirow{4}{*}{\cite{smith2017analyzing,graziano2021establishment,hannah2021cyber,smith2020view,smith2020understanding}} & Receive, parse, and collect messages with a radio receiver& Theft of aircraft statistics  & AC5, AC6& No Confidentiality& Information disclosure   
\\ \cline{2-2}\cline{4-8}
  & Ghost aircraft &  & Respond to interrogations to maintain false track and declare a threat& \multirow{3}{*}{Loss of safety}& \multirow{3}{*}{\makecell[l]{AC3\\(AC6\textbar{}AC12)\\AC18}} & No Authentication& Spoofing
\\ \cline{2-2}\cline{4-4}\cline{7-8}
  & Alpha-beta drop&  & Issue corrupted response on behalf~of targeted aircraft to damage its reliability & &   & No Availability   & DoS
\\ \cline{2-2}\cline{4-4}\cline{7-8}
  & Address flooding&  & Craft and inject multiple aircraft acquisition squitters with unique ICAO address numbers& &   & \multirow{2}{*}{No Authentication} & \multirow{2}{*}{Spoofing}
\\ \Xhline{2pt}
Engine Alerting& Engine data manipulation   & \cite{duchamp2016cyber,EICASKaspersky}   & Inject malicious engine alerting messages& Loss of safety& AC7, AC15&&  
\\ \Xhline{2pt}
\multirow{3}{*}{IFEC}& Display messages on panels & \multirow{3}{*}{\cite{IFECIOACTIVE,IFECsystem}}   & Gain control of passenger's display systems& \multirow{2}{*}{Sow fear}& \multirow{3}{*}{\makecell[l]{AC3, AC7,\\AC15\\(AC21\textbar{}AC24\\\textbar{} AC25)}}   & No Integrity& Tampering
\\ \cline{2-2}\cline{4-4}\cline{7-8}
  & Control cabin lightning&  & Gain control of cabin attendant panel& &   & No Authorization & Elevation of privileges   
\\ \cline{2-2}\cline{4-5}\cline{7-8}
  & Access passenger's credit card details   &  & Collect information from IFEC servers   & Theft of passengers' information&   & No Confidentiality& Information disclosure   
\\ \Xhline{2pt}
\multirow{5}{*}{EFB}& Remote control & \multirow{5}{*}{\cite{EFBReport,EFBPPT,vanhoef2017key}} & Exploit Bluetooth, Wi-Fi, or Cellular link  & Loss of control   & (AC22 \textbar{} AC23 \textbar{} AC24)& No Integrity& Tampering
\\ \cline{2-2}\cline{4-8}
  & Retrieve manuals~and technical logs&  & Collect stored information on applications& Theft of aircraft statistics  & \multirow{4}{*}{\makecell[l]{AC16,\\(AC22 \textbar{} AC23\\\textbar{} AC24)}} & No Confidentiality& Information disclosure   
\\ \cline{2-2}\cline{4-5}\cline{7-8}
  & Modify flight planning~and navigation data   &  & Hack into hosted applications and control~their actions, modify navigation data base information&
  \multirow{2}{*}{\makecell[l]{Manipulation of\\ control}}  &   & No Authorization & Elevation~of~privilege   
\\ \cline{2-2}\cline{4-4}\cline{7-8}
  & Install malicious application&  & Replication through removable media/social engineering/malicious device & &   & No Integrity& Tampering
\\ \cline{2-2}\cline{4-5}\cline{7-8}
  & Control cabin lightning&  & Abuse hosted applications and~modify their storage and responses& Sow fear&   & No Confidentiality& Information disclosure  
\\ \Xhline{2pt}
\end{longtable}}
\end{landscape}

\section{\label{sec:taxonomy}Extending MITRE Framework for Aviation}
In order to standardize the knowledge on and understanding of threats to cyber security in the aviation field, we followed the ATT\&CK (Adversarial Tactics, Techniques, and Common Knowledge) model utilized by MITRE~\cite{MITRE} to classify attack tactics, techniques, sub-techniques, and procedures.
The MITRE model systematically explains the adversaries' actions to be executed in the target device/system/domain from an adversary's perspective.
MITRE ATT\&CK covers the enterprise, mobile, and industrial control system fields.
In this section, we divide existing and possible actions in the aviation domain into a variety of tactics, techniques, and sub-techniques.

\begin{itemize}[leftmargin=*]
  \item \textbf{Tactics} represent the \textit{"why"} of the technique: the  the adversary's tactical goals during an attack.
  
  \item \textbf{Techniques} represent \textit{"how"} an adversary achieves a tactical objective by performing an action.
  
  \item  \textbf{Sub-Techniques} are more specific descriptions of the adversarial behavior used to achieve a goal, while techniques represent the broad actions an adversary takes to achieve a tactical goal, 
  
  \item  \textbf{Procedures} procedures are the specific implementation the adversary uses for techniques or sub-techniques. In this work, a procedure is a reference to the implementation of a technique as applied in academic studies or in industrial applications.
\end{itemize}

\begin{figure*}[ht]
    \centering
    \includegraphics[width=1\textwidth]{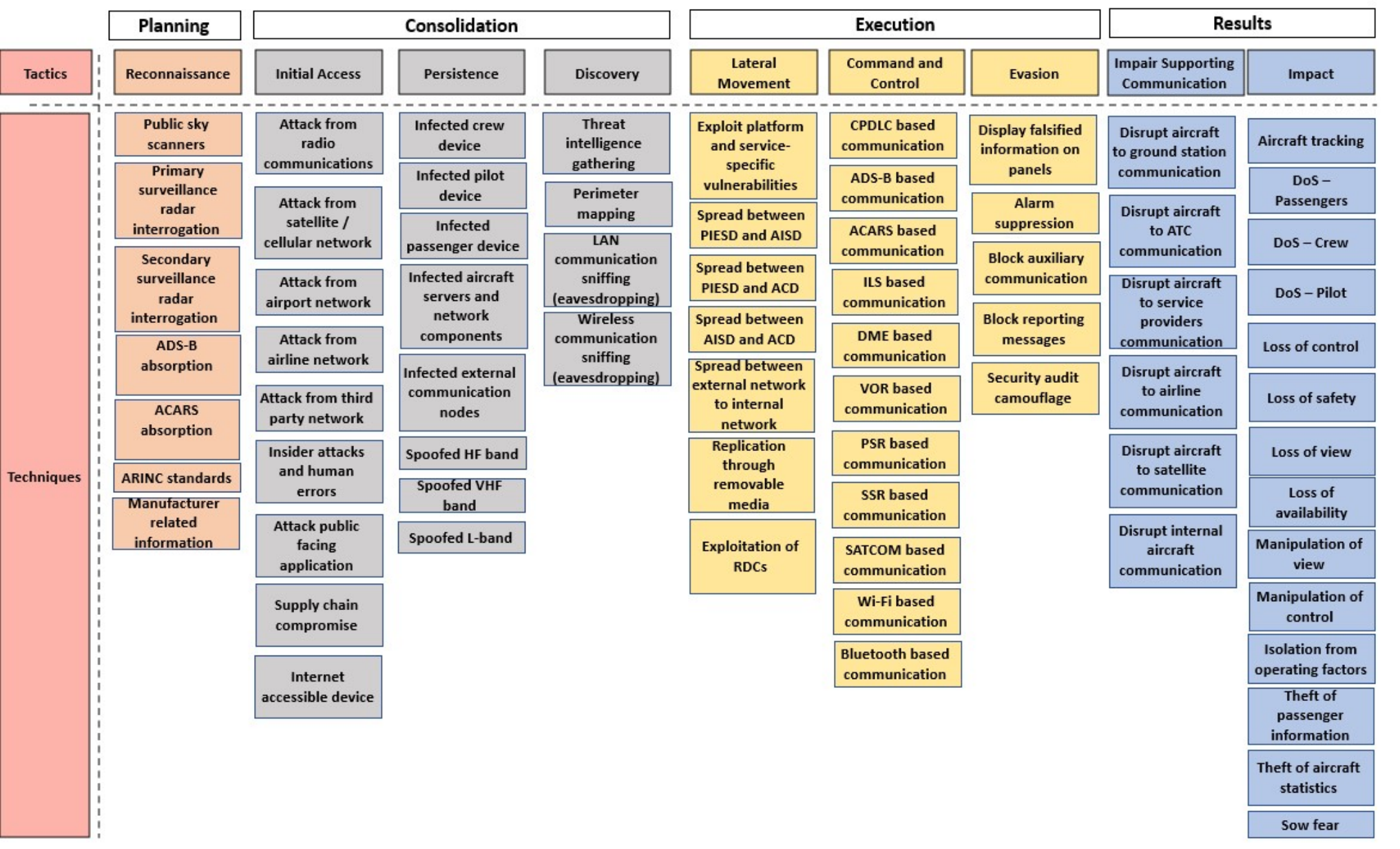}
    \caption{Taxonomy - aviation matrix tactics and techniques.}
    \label{fig:Taxonomy}
\end{figure*}

We opted to align our taxonomy with the ATT\&CK model, because the attack sequence diagram can indicate the adversary’s behavior and capabilities, limitations on how adversaries (or a specific group/APT) can compromise the system, and the loosely protected systems and connections that require more rigorous security. 
Moreover, the matrix representation helped us build a systematic categorization and taxonomy based on the known attacks and retained the attack phases as a sequence chart.
The matrices in Figure~\ref{fig:Taxonomy} represent the various tactics in the aviation field (columns) and the techniques used to achieve them (the individual cells).

\subsection{Reconnaissance}
Using the reconnaissance tactic, the adversary tries to gather information, and identify and select targets.
The adversary collects information that can be used to support the targeting and selection of attack techniques in other phases of the attack lifecycle (i.e., locating potential target systems/networks for initial access).
The adversary can use multiple techniques:

\noindent\textbf{Public sky scanners}: The adversary can search for available information using online flight trackers (i.e., OpenSky\footnote{https://opensky-network.org/}, Flightradar24\footnote{https://www.flightradar24.com/}). 
Information about flight plans, both visual flight rules (VFRs) and instrument flight rules (IFRs), can also be found online using services such as SkyVector.\footnote{https://skyvector.com/}
\newline\textbf{PSR interrogation}:
PSR is the only surveillance sensor used in civil aviation that does not require any on-board equipment to locate aircraft, therefore an attacker can use PSR equipment to gather information about the aircraft's location without any dependence on its components.
\newline\textbf{SSR interrogation}:The ability to use SSR interrogation depends on the components installed on the aircraft, as the SSR relies on targets equipped with a radar transponder.
Civil aircraft may be equipped with transponders capable of operating in different modes:
Mode A  equipment only transmits an identifying code. Mode C equipment automatically obtains the aircraft altitude or flight level, and Mode S equipment has altitude capability and enables data exchange.
\newline\textbf{ADS-B absorption}:
ADS-B is a surveillance technique that relies on aircraft or airport vehicles broadcasting their identity, position, and other information derived from onboard systems (e.g., GNSS). 
An adversary can capture broadcast ADS-B messages using an ADS-B IN receiver and obtain information about an aircraft's GPS location, altitude, ground speed, and more.
\newline\textbf{ACARS absorption}:
ACARS data processing can provide an adversary with extensive information about air traffic control, aeronautical operational management, and airline administrative control, e.g., route updates, weather updates, and even information about special passengers ("housekeeping services").
\newline\textbf{ARINC standards}: The adversary can gather supported ARINC standards on avionics, cabin systems, protocols, and interfaces provided by Rockwell Collins.\footnote{https://www.rockwellcollins.com/}
\newline\textbf {Manufacturer related information}: In order to obtain supporting information that can be leveraged by the adversary to aid in other attack phases. Collecting information regarding relevant devices implemented onboard is crucial and can be found on different forums and sites, i.e., the FAA engineering database~\cite{FAADATABASE}, specification forums~\cite{specifications}, manufacturer specifications\footnote{https://modernairliners.com/} and patents (e.g., Boeing's patent for the e-Enablement network implementation~\cite{BoeingPatent}).
      
\subsection{Initial Access}
With the initial access tactic, the adversary tries to establish an attack vector by gaining access to the targeted system/environment.
By utilizing all the means of access into the system from the outside world and public network,
there might be various entry vectors for gaining an initial foothold within an aviation system.
\newline\textbf{Attack via radio communication}: Radio communication refers to ground-based communication using HF and VHF, to communicate with the aircraft. Utilizing HF, VHF, and L-band communication to enter the network is convenient for the attacker, since, by using simple COTS, he/she can transmit at different ranges to the various aircraft systems (ILS, ACARs, VOR, DME, ADS-B, and more).
At the same time, using these means of communication requires the attacker to have a line of sight to the target being attacked. 
\newline\textbf{Attack from satellite/cellular network}:
The aircraft systems use satellites for communication and navigation.
For example, Satellite Voice-equipped aircraft can initiate calls using Inmarsat\footnote{https://www.inmarsat.com/} or Iridium\footnote{https://www.iridium.com/} assigned security phone numbers, and IFEC in-flight Wi-Fi uses a satellite-based Wi-Fi system (e.g., Viasat\footnote{https://www.viasat.com/enterprise-and-mobility/aviation/commercial/}).
In addition, air traffic management systems, such as the ADS-B and CPDLC use satellite data links. To this end, satellite range control can provide a diverse attack surface for PIESD and ACD networks while controlling a wide geographical area.
\newline\textbf{Attack from airport network}: Aircraft connect to the airports' wireless networks for maintenance; both wireless and cellular connections can be compromised by adversaries. 
Examples of potential devices that can be exploited using wireless vulnerabilities (e.g., KRACK vulnerability~\cite{feher2018effects}) include standalone EFB tablet devices, the aircraft terminal wireless LAN unit (TWLU) used for airport gatelink connectivity, the connectivity and crew wireless LAN unit (CWLU) used for maintenance laptop connectivity, and the wireless dataloaders used for software updates.
\newline\textbf{Attack from airline network}:
By penetrating the airline network, an attacker can achieve wide access to the aircraft it operates while interfering with control, operation, and maintenance processes. 
In addition, an attacker can tamper and interfere with  with pilot communication using systems such as VHF voice CPDLC, and ACARS.
\newline\textbf{Attack from third-party networks}:
Many services have access to various systems and different functions in the aircraft. 
These services can be used to bypass aircraft systems. 
Examples of possible targeted services include maintenance services (e.g., ACT services\footnote{https://www.actservices.de/}), management services (e.g., Teledyne), flight-planning services (e.g., Honeywell, Lido), support services (e.g., AMETEC), and development services (e.g., Keysight).
\newline\textbf{Insider attacks and human errors}: Insider attacks involve both intentional attacks and unintentional mistakes by a human with access to any component of the avionics ecosystem. An insider can be a crew member, maintenance worker, or any other airport/airline employee.
Human errors and nsider assistance can be used to bypass various security measures and gabin physical access to components, and they are often leveraged by adversaries as initial access techniques. 
\newline\textbf{Public-facing application}: The adversary can exploit a public-facing application that does not require special access privileges; such applications can be used to seek and obtain access points to aircraft systems. Examples of potential surfaces are the EFB application server/store, administration websites, and airline websites.
\newline\textbf{Supply chain compromise}: Supply chains can be used to gain access to platforms around the world. 
Devices and software can become compromised if an adversary tampers with the manufacturing process of a product by installing a rootkit or hardware-based spying component. 
The targeted devices include various sensors, remote data concentrators (RDCs), actuators, cabin devices (e.g., IFEC cell modems, SATCOM modems, and smart monitors).
\newline\textbf{Internet accessible devices}: An adversary can infect Internet-accessible devices such as an IFEC content server or maintenance laptop. 
More sophisticated vectors can target SATCOM antennas, as illustrated by Santamarta \textit{et al.}~\cite{santamarta2014satcom}.

\subsection{Persistence}
The persistence tactic aims to allow continuous access to aviation systems. 
In order to maintain access in the face of en-route and survive restarts, credential changes, and other interruptions, there are a variety of actions an adversary can perform, ranging from changing settings to interfering with system files or hardware. \\
\textbf{Infected crew device}: An adversary can gain access to the aircraft crew's terminals or their personal devices (e.g., mobile phones) thus ensuring consistent proximity to the aircraft. \\
\textbf{Infected pilot device}: Infecting pilot-owned devices, such as the EFB tablet, ensures network access to the aircraft's cockpit and core systems and an understanding of the activities of the pilot's activities during the flight. \\
\textbf{Infected passenger device}: Attacking passenger devices allows persistence but only for the duration of the passenger's flight.
Connectivity to a passenger's device may allow connectivity to the cabin systems, particularly the IFEC system and the aircraft's Wi-Fi network. \\
\textbf{Infected aircraft servers and network components}: In order to gain access to the aircraft servers and network components, adversaries can use different techniques seen in the ATT\&CK enterprise taxonomy, for example:

\begin{itemize}[leftmargin=*]
    \item \textit{modify configurations -} An adversary can gain access to a system by editing the configuration file that defines its features.  Examples of configuration files in the aircraft core systems include loadable media CONFIG.LDR and EXCONFIG.LDR files, operation program configuration (OPC) files, and airline modifiable information (AMI) files.
    \item \textit{modify programs and applications -} An adversary can modify a program in order to affect the way it interacts with other systems and devices. Modified applications can be used to add new logic that enables persistence on the host device.
    For example, the EFB device has many applications (e.g., weight and balance applications, flight planning applications, and performance applications) that may serve as an attack surface.
    \item \textit{System firmware -} Device firmware updates can be delegated using a software update package provided remotely~\cite{IATALOADABLE}.
    \item \textit{Module firmware -} Device firmware such as software-loadables, LRUs and other modular hardware devices can be install or modified to achieve persistence and provide secret access points.
\end{itemize} 
\textbf{Infected external communication nodes}: An adversary can own an allegedly authorized datalink by penetrating trusted remote service networks such as the CPDLC provider's network, including the ATN network (VHF data link operated by ARINC and SITA) and FANS network (satellite communications provided by Inmarsat). Moreover, an adversary can launch rogue endpoints, e.g., rogue GSM towers that communicate with the IFEC system's cell modem or a ground station that transmits FIS-B and TIS-B messages. \\
\textbf{Infected HF band}: By spoofing over the HF communication range an attacker can maintain contact with an aircraft at short range. \\
\textbf{Infected VHF band}: By spoofing over the VHF communication range, an adversary can operate and affect various systems (e.g., the ILS, DME, and VOR systems)while maintaining a line of sight to the attacked aircraft. \\
\textbf{Infected L-band}: By spoofing over the UHF and its upper bound (L-band)  communication range, an adversary can operate and affect satellite-based systems (e.g., the TCAS, ACAS X, and ADS-B systems) from far away.

\subsection{Discovery}
In order to gain knowledge on the environment, adversaries use the discovery tactic.
This tactic consists of a collection of techniques designed to allow an attacker to determine the options for advancing, what measures to take, and how to spread within the network. \\
\textbf{Threat intelligence gathering}:
An adversary can use dedicated search engines (e.g., Shodan\footnote{https://www.shodan.io/}  and Censys\footnote{https://censys.io/}) that gather information about vulnerable devices and networks, in order to identify exposed critical nodes that are sometimes visible on public networks due to misconfigurations. \\
\textbf{Perimeter mapping}:  Devices' communication patterns can be discovered using connection enumeration. 
An adversary can use different tools to determine the role of a device on the network and identify its connections to other systems. 
Moreover, adversaries may attempt to obtain a list of different systems and components using network identifiers (e.g., MAC addresses, TTL values). 
By obtaining IP addresses and identifying the type of operating system, an adversary can choose how and where to spread. \\
\textbf{LAN communication sniffing (eavesdropping)}: Adversaries can use sniffing tools in order to monitor or capture information transmitted through the network, e.g., eavesdrop file transfer services traffic for discovering data and credentials. 
Sniffing can also be useful to detect the current aircraft state by capturing the OOOI events from sensors and gateways. \\
\textbf{Wireless communication sniffing (eavesdropping)}: Adversaries can use the wireless range to obtain location signals, i.e., Mode-S (1030/1090 MHz frequency) or report ACARS signals at a frequency of 131.550 MHz.

\subsection{Lateral Movement}
The lateral movement tactic consists of techniques used by adversaries to spread between components in the aircraft.
An adversary might want to move between different domains to gain access to more critical systems, e.g., move from the PIESD to the AISD or ACD.
This tactic is usually applied after performing discovery techniques and identifying a target destination.

\noindent\textbf{Exploit platform and service-specific vulnerabilities}:
While many components within the aircraft run popular real-time operating systems (e.g., VxWorks, QNX, ThreadX), significant weaknesses are revealed in some of these systems. For example, the Armis Security team has already identified 11 vulnerabilities (URGENT/11)~\cite{seri2019critical} in the VxWorks OS kernel, e.g., an opcode stack overflow that can lead to arbitrary code execution.
In order to abuse an existing platform, an adversary can use different utilities in order to inject hooks or abuse API (e.g., Frida\footnote{https://frida.re/}). \\
\textbf{Spread between the PIESD and AISD}: 
As described above in Subsection~\ref{PIESDAISD}, there are overlapping connections between both networks (PIESD and AISD). 
As the AISD contains information services systems, the attacker can use the AISD network to communicate with third-party providers for accessibility and information leakage, or as part of further lateral movement to the ACD network. \\
\textbf{Spread between the PIESD and ACD}:
The passengers' related networks and the control systems should be separated. However, in practice, there are are overlapping connections between the networks (e.g., the CSS has dual connections: a connection to the in-flight entertainment system and a link to the ACD switch designed for transferring audio between the pilot and the cabin. \\
\textbf{Spread between the AISD and ACD}:
As described above in Subsection~\ref{AISDACD}, there are overlapping connections between the networks for non-essential applications (e.g., the cabin services system and flight deck recorder) ued to connect the cockpit systems to other aircraft systems. In this way, an adversary that reaches the ACD network can access the core air traffic, information, and navigation systems. \\
\textbf{Spread between the external network and internal network}:
As described above in Subsection~\ref{EXTERNALCONNECTION}, the aircraft serves as a flying domain controller; therefore, it depends on communication between the ground, its surroundings, and satellites. Thus an attacker can find many access points from an external network which is accessible from the Internet into an internal network of the aircraft. \\
\textbf{Replication through removable media}: In order to achieve access to hardened components/networks  (e.g., air-gapped devices), adversaries may opt to choose the replication through removable media technique. Aircraft contain different systems with portable USB connections (e.g., cabin panels, smart screens, EFB tablets, and IFEC crew terminals). Shellcode and payloads can be activated when devices are plugged in, via AutoRun~\cite{thomas2009rise}. \\
\textbf{Exploitation of RDCs}: An adversary can abuse RDCs to access controllers, systems, and actuators. For example, controlling the actuator control electronics unit (ACE) will provide the adversary with direct control of the flight control surface, by obtaining all inputs and communicating with the flight computer. 

\subsection{Command and Control}
After initial access to one of the aircraft systems has been obtained, and various types of utilities (tools and malware) are distributed  between the aircraft components, adversaries must establish a communication channel to command and control the scattered assets. 
In order to create a stable method of communication, the attacker will usually try to find a stealthy or seemingly legitimate method of communication.
To achieve this goal, an adversary can abuse communication methods embedded in the aircraft's systems (Generalized as \textit{Standard Protocol/Datalink Misuse}).
To apply this technique, an adversary can utilize aircraft systems and protocols that communicate with the outside world (e.g., satellites, ground stations, airports, airlines, service providers) to receive or transmit the information.
The following communication methods between air to the ground are examples of the potential channel that an adversary can use for the use of command and control: 
\newline \textbf{VHF/HF/SATCOM based communication} - Used by the ACD as part of the flight and embedded control systems and cabin systems (e.g., ADS-B, CPDLC, ACARS, and more).
\newline \textbf{Wireless LAN based communication} - Used by the AISD as part of administrative support, flight support, and maintenance (e.g., communication with airport network).
\newline \textbf{Broadband/Cellular based communication} - Used by the PIESD as part of the in-flight entertainment and passenger internet systems.
All these means of communication can allow an adversary to transmit and receive data via different ranges and accuracy.
\subsection{Evasion}
The evasion tactic consists of techniques that adversaries use to avoid detection once they have gained access to the system.
Usually this tactic's techniques require active steps to conceal the adversary's presence or to remove evidence. \\
\textbf{Display falsified information on panels}:  Adversaries can modify the content of different panels (cockpit control panels, crew terminal panels, EFB applications, graphical interfaces) to disrupt the crew and pilot's behavior or evade detection.\\
\textbf{Suppress alarms}: Adversaries may manipulate failure alert systems (e.g., collision or engine failure alert systems) to avoid the detection of system damage. \\
\textbf{Block auxiliary communications}: Adversaries can block the CPDLC in order to disable communication between an aircraft and the ground station. \\
\textbf{Block sensor data}: Adversaries can block sensor or RDC communication  to interfere with signals that indicate the aircraft state. \\
\textbf{Spoof/block reporting messages}: Adversaries can spoof and block ACARS signals, OOOI events (engine performance, monitoring, fault reports, fuel status reports, selective calls, passenger services, maintenance reports, and other information such as load and balance) to hide actions that affect the aircraft  state~\cite{duchamp2016cyber,zhang2017analysis}. \\
\textbf{Security audit camouflage}: Security audit camouflage refers to any means by which an adversary can remain undetected from audit measures within the systems and software. 
The following methods are borrowed from the ATT\&CK enterprise framework, as currently there is no evidence for audit camouflage in avionics systems, but at the same time the actions required for hiding information are similar:

\begin{itemize}[leftmargin=*]
\item\textit{Masquerading}: Adversaries can use masquerading techniques to disguise a malicious application or executable as another file (e.g., log files, config files).

\item\textit{Indicator removal on host}: Adversaries may delete or alter generated artifacts on a host system, including logs or captured files, in their efforts to cover their tracks.

\item\textit{Rootkit}: Rootkits are programs that hide their existence and the presence of malware by intercepting the operating system's API calls that supply relevant information. 
Adversaries can use rootkits in order to hide their malicious tools and payloads.
\item\textit{Exploitation for evasion}: Adversaries can exploit a software vulnerability to take advantage of an error in a program or service, or within the operating system software or kernel itself, to evade detection.
\end{itemize}

\vspace{-5pt}
\subsection{Impair Supporting Communication}
This tactic is an additional to those proposed by ATT\&CK. Avionic systems consists of various communication, navigation, display and management systems integrated in aircraft to perform individual functions. The proper functioning of these systems depends on continuous operation and collaboration with external services (e.g., ground stations, satellites, airports, airlines, operators). As part of a multi-stage attack vector, one of the attacker's goals will likely be to disrupt such collaboration and the supporting systems.
Some of the collaborations an adversary can disrupt are presented below:
\newline \textbf{Between aircraft and ground station - } Disrupting ground-based and navigation communication aids has an impact on the navigation systems, e.g., the DME, VOR, NDB, and ILS.
\newline \textbf{Between aircraft and ATC - } Disrupting communication with ATC ground stations usually refers to VHF and L-band spoofing or jamming, e.g., interference with ADS-B, FIS-B, TIS-b, SSR interrogations. Disruption of these collaborations is of great significance as it affects the perception of the pilot, crew, and operators regarding the aerial state.
\newline \textbf{Between aircraft and service providers - } Disrupting communication with third-parties providers (e.g., Inmarsat, Iridium, Viasat, Teledyne, Honeywell, AMETEC, Keysight). This interference has a great impact on the aircraft's ability to communicate with the outside world, as satellite services are deployed by external companies.
\newline \textbf{Between aircraft and airlines - } Disrupting communication channels with airlines, e.g., Airline Operational Control (AOC) and Airline Administrative Control (AAC) messages.
\newline\textbf{Between aircraft and satellites - } Disrupting satellite communications (SATCOM). This action can be caused by an adversary who uses powerful transmitter to beam a jamming message towards the satellite. 
\newline \textbf{Internal aircraft connections - } Disrupting the communications between internal systems and the network that transfer information between them (e.g., by harming the AISD or PIESD switches).

\subsection{Impact}
As described in Section~\ref{sec:AdversariesThreat}, adversaries have different goals regarding impact. 
The impact tactic consists of techniques the adversary uses to disrupt, compromise, destroy, and manipulate the integrity and availability of an aircraft's components and systems.
While the effect some impact techniques have on those on the aircraft is less obvious, such techniques can directly affect the privacy or personal safety of those onboard.
\newline \textbf{Aircraft tracking:} Adversaris can obtain the exact location of the aircraft and follow its route.
\newline \textbf{Passenger DoS}: Adversaries can control cabin operations (e.g., the passenger panels, Wi-Fi, and content server).
\newline \textbf{Dos - Crew }: Adversaries can control the IFEC crew terminal, interfering with the cabin crew's ability to control the IFEC system.
\newline \textbf{DoS - Pilot}: Adversaries can take control of the cockpit control panels and EFB device to interfere with the pilot's ability to make decisions and apply them.
\newline \textbf{Loss of control}: Adversaries can obtain control of different sensors and actuators, which are crucial for the aircraft's functionality. For example, gaining access to the ACE component can be used to achieve full control of actuators and RDCs .
\newline \textbf{Loss of safety}: Adversaries can cause dangerous situations that affect the safety of the passengers and crew, such as cause landing failure (spoofed/jammed ILS signals~\cite{sathaye2019wireless} or GPS signals~\cite{miralles2020assessment,kovzovic2021spoofing}), trigger false TCAS alerts or harm the TCAS system~\cite{ILSandTCASSpooging, hannah2020traffic,hannah2021cyber}, manipulate the engine alerting system or navigation system, or interfere with the ADS-B surveillance system~\cite{costin2012ghost}.
\newline \textbf{Loss of availability}: Adversaries may attempt to disrupt essential components or systems to prevent the proper transfer of information by interfering with the channels used to communicate  with ground stations (e.g., SATCOM/HF/VHF).
\newline \textbf{Manipulation of view}: Adversaries can interfere with the pilot's view, causing a sustained or permanent loss of view, by compromising the flight deck instrument display system (EFIS). 
The EFIS normally consists of a primary flight display (PFD), multi-function display (MFD), and an engine indicating display.
\newline \textbf{Manipulation of control}: Adversaries may manipulate the set point values and parameters, such as the intermediate waypoints in the navigation database. Since waypoints can be used to change routes, modifying them may affect the flight route. 
\newline \textbf{Isolation from operating factors}: Adversaries may aim to interfere with the pilot's and flight crew's ability to receive messages from ground stations and operators in order to ensure their complete isolation and eliminate their ability to receive guidance or assistance.
\newline \textbf{Theft of passenger information}: Adversaries can exploit IFEC systems to steal passengers' data (i.e., credit cards details, passport  numbers).
\newline \textbf{Theft of aircraft statistics}: Access to information stored in the aircraft and collected from the various sensors may be used by adversaries to obtain information about the aircraft's activity and be used for a variety of purposes, such as industrial espionage.
 \newline \textbf{Sow fear}: Adversaries can take control of the aircraft's smart monitors and display messages.

\section{Demonstration - Modeling the Taxonomy Framework \label{sec:usecases}}
In this section, we model two attacks according to our proposed taxonomy described in Section~\ref{sec:taxonomy} and demonstrate how the two examples can be analyzed using the proposed tactics and techniques.

\vspace{-5pt}
\subsection{Case Study - Hacking into the Flight Management System}
\begin{figure}[h!]
    \centering
    \includegraphics[width=1\linewidth]{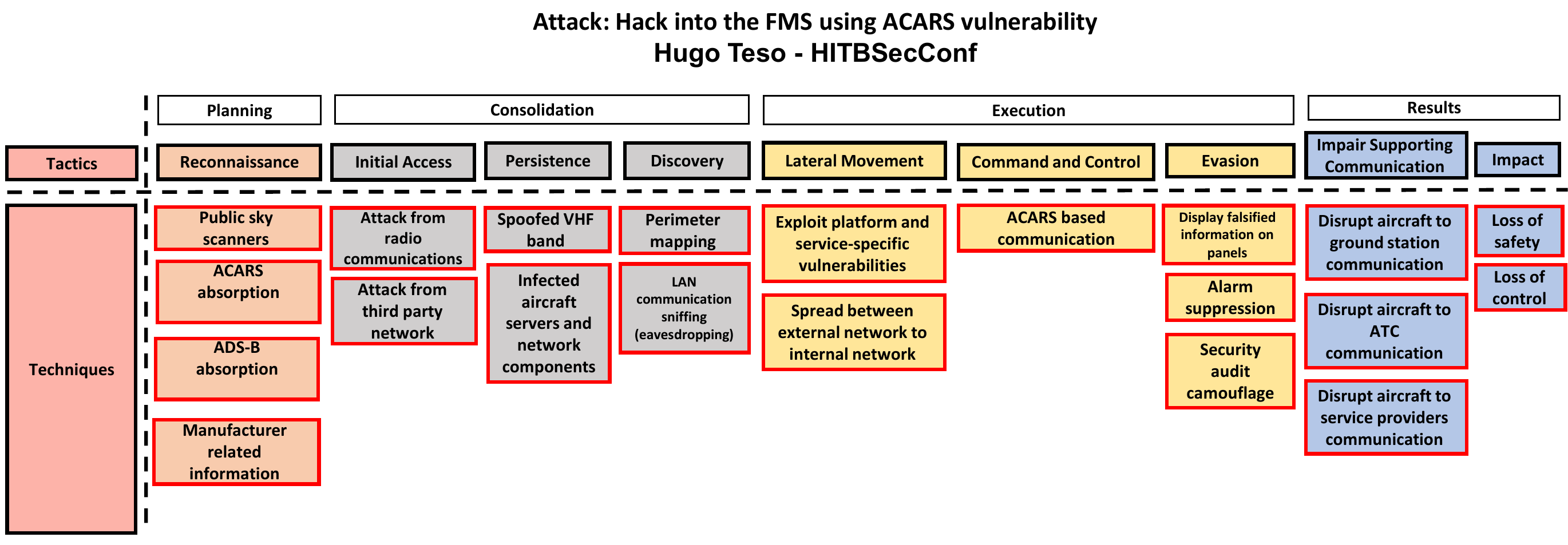}
    \caption{Exploiting the FMS using ACARS vulnerabilities - modeling taxonomy.}
    \label{fig:HUGOMATRIX}
\end{figure}
Hugo Teso demonstrated the following case study at HITBSecConf~\cite{teso2013aircraft} --- hacking into the FMS using ACARS vulnerability. His experiment was conducted under laboratory conditions using original components.
The attack targeted the onboard FMS component by exploiting its weakness via the ACARS datalink. Information regarding the target aircraft was collected using the ADS-B protocol. 
SIMON is a two-way communication exploitation agent used to inject various payloads and plugins. 
\newline\textbf{Assumptions:} 
\begin{itemize}
    \item[-] The adversary is intentionally trying to influence the FMS.
    \item[-] The target flight is within the adversary's line-of-sight.
    \item[-] The adversary is located on the ground.
\end{itemize}
We now describing the attack vector step by step following the taxonomy.
\begin{itemize}[wide, labelwidth=!, labelindent=0pt]
    \item \textit{\textbf{Reconnaissance - }} 
    Hugo used the ADS-B protocol \emph{(ADS-B absorption)} to obtain information about the aircraft in the area; the data was obtained using a receiver and \emph{public sky scanners} (e.g., Flightradar24, OpenSky Network).
    Additional information such as flight plans and route clearance information were obtained using \emph{ACARS absorption}.
    Finally, to perform this research, construct his lab, and gather accurate data, Hugo acquired the Honeywell FMS and Teledyne ACARS from eBay\footnote{https://www.ebay.com/}; therefore he was able to obtain \emph{manufacturer-related information}.
    \item \textit{\textbf{Initial Access - }} 
    Since the attacker utilizes SDR to craft and transmit ACARS signals towards the targeted aircraft, we can conclude that the initial access is achieved through the \emph{radio access network}. Furthermore, Hugo mentioned that by using an SDR, he was able to target nearby aircraft and gain control of \emph{third-party networks} that provide ACARS services (i.e., SITA and ARINC),
    \item \textit{\textbf{Persistence - }} With regard to persistence, Hugo kept transmitting ACARS messages and \emph{spoofed the VHF band} in order to continue  with the attack immediately after installing the initial payload with an exploitation agent (denoted as SIMON).
    \item \textit{\textbf{Discovery - }} After installing SIMON, Hugo obtained two-way link communication with the aircraft, which enabled him to use \emph{perimeter mapping} and \emph{LAN communication sniffing} to find connected components that may be vulnerable.
    \item \textit{\textbf{Lateral Movement - }} The first movement from the ground to the aircraft took place by spreading from the external network to the internal network (e.g. communicate through providers network with components within the ACD network).
    Further movement requires the \emph{exploitation of platforms and service-specific vulnerabilities} (e.g., ACARS vulnerabilities).
    In order to perform these exploitations, Hugo used reverse engineering tools and studied real-time OSs that the FMS deploys (VxWorks and DeOS). 
    \item \textit{\textbf{Command and Control - }} In order to communicate with SIMON, Hugo set up a two-way \emph{communication link over ACARS}. By doing so, he could upload new plugins and script. The only constraint is that each ACARS message can only contain up to 220 characters and the ACARS management unit can transmit 16 subsequent messages at a time.
    \item \textit{\textbf{Evasion - }} As SIMON is not designed as a rootkit, \emph{alarm suppression} and \emph{security audit camouflage} are necessary in order to keep the autopilot active so the attacker remains stealthy and in control.
    \item \textit{\textbf{Impair Supporting Communications - }} As Hugo interfere  with the ACARS communication, he managed to disrupt the communication between the aircraft and \emph{ATCs, ground stations, and service providers}.
    \item \textit{\textbf{Impact - }} By controlling the FMS, an adversary can communicate with engine and fuel systems, navigation receivers, surveillance systems, and flight controls.
    Therefore, tampering with the data -n the FMS or damaging it has destructive potential, threatening \emph{aircraft safety} and \emph{aircraft control}.
\end{itemize}

\subsection{Case Study - Spoofing Ghost Aircraft using ADS-B}

\begin{figure*}[h]
    \centering
    \includegraphics[width=1\linewidth]{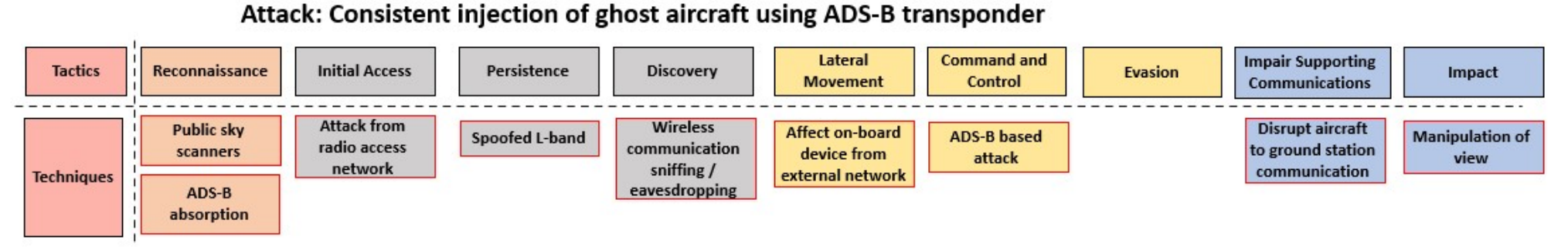}
    \caption{Consistent injection of ghost aircraft - modeling taxonomy.}
    \label{fig:USECASE1}
\end{figure*}
The ADS-B system is prone to spoofing, flooding, tampering and eavesdropping attacks as described at various works (e.g.,~\cite{viveros2016analysis,costin2012ghost,mirzaei2019security,eskilsson2020demonstrating}).
\newline\textbf{Assumptions:} 
\begin{itemize}
    \item[-] The adversary intentionally tries to influence the information obtained on a particular flight.
    \item[-] The target flight is in the adversary line-of-sight.
    \item[-] The adversary located on the ground.
\end{itemize}
We now model the ghost-aircraft attack vector step by step following the taxonomy.
The taxonomy matrix applied on this case study is illustrated in Figure~\ref{fig:USECASE1}.
\begin{itemize}[wide, labelwidth=!, labelindent=0pt]
    \item \textit{\textbf{Reconnaissance - }} To obtain information about the latest aerial state, the aircraft positions, exact coordinates, and altitude of the targeted aircraft, an adversary needs the ability to analyze the ADS-B signals \emph{ (ADS-B absorption)} using the ADS-B IN receiver or \emph{public sky scanners} (e.g., Flightradar24, OpenSky Network).
    \item \textit{\textbf{Initial Access - }} Since the attacker utilizes the ADS-B OUT transmitter/Mode S transmitter to transmit ADS-B signals towards the targeted aircraft, it can be concluded that initial access is achieved through the \emph{radio access network} or the L-band.
    \item \textit{\textbf{Persistence - }} For persistence, the adversary has to continuously \emph{spoof the L-band} with ADS-B messages in the range of absorption of the ADS-B IN receiver on the target.
    \item \textit{\textbf{Discovery - }} Although the adversary does not have to advance to any network, to get an idea of his/her broadcasts' visibility, the adversary can process his/her signals using a receiver and view their appearance. To do so, the adversary has to \emph{eavesdrop on the L-band}.
    In addition, the adversary must monitor air traffic and ensure that his/her broadcasts correspond with the flight paths in the area and nearby aircraft. 
    \item \textit{\textbf{Lateral Movement - }} The adversary affects the information spread in the internal systems of the aircraft while communicating with the aircraft from the outside; therefore the adversary \emph{affects onboard devices from an external network}.
    \item \textit{\textbf{Command and Control - }} The adversary abuses the standard protocol (ADS-B) by transmitting messages that meet the requirements of the protocol in the appropriate range. Therefore the attack is an \emph{ADS-B based attack}.
    \item \textit{\textbf{Evasion - }} The attack does not include an evasion attempt.
    \item \textit{\textbf{Impair Supporting Communications - }} As the adversary is
    on the ground, there is a possibility that the malicious transmissions will be received by ADS-B ground stations; this would allow the attacker to disrupt the channel \emph{between the ground station and the aircraft}. 
    \item \textit{\textbf{Impact - }} As the adversary tampers with the data displayed on the aircraft and ground station panels, he/she is therefore \emph{manipulating the view}.
\end{itemize}

\section{\label{sec:Gaps}Scientific Gaps and Research Directions}
This section deals with the scientific gaps and subjects requiring further investigation identified in our work. We also discuss issues arising from the lack of complete and reliable information available in the field of offensive cyber security and suggest directions for further research.
\begin{enumerate}[wide, labelwidth=!, labelindent=0pt]
\item \textbf{Verification of Questionable Information}
Research on offensive cyber security, especially when critical systems are involved, is challenging, since in many cases the complete details and data on an attack method cannot be disclosed or published due to censorship, security concerns, and other constraints. 
Therefore, it is important to note that some of the attacks mentioned in the paper were presented at various conferences without sufficient evidence. 
In addition, there are several cases where the veracity of claims is questionable, e.g., avionics experts dispute a security researcher's claim that they accessed in-flight entertainment and flight systems from their seat on an aircraft~\cite{ChrisRoberts}. 
Therefore, in order to confirm some of the attack vectors presented, additional experiments should be conducted.
\item \textbf{Scientific Gaps}
In the field of avionics, there are several scientific gaps and areas where there is insufficient knowledge and understanding. 
In the case of avionics, this lack of understanding poses a real danger, and there is a need for further research to reduce threats to aircraft and those aboard them.
\begin{itemize}
\item \textbf{Protection Methods}
Many studies have addressed the need for protection against attacks on specific systems. The solutions proposed generally require redesigning the system or its components, adding components or sensors, etc.
However, there is a need for a solution that takes a wide view of an aircraft as a broad and uniform computerized platform with many intrusion surfaces. The practice of utilizing many systems on an attack vector has hardly been examined. In this paper, we show how different systems and components can be used throughout the stages of attacks to achieve the adversary's desired impact.
\item \textbf{Adoption and Adaptation of Existing Solutions}
The computer systems on an aircraft include real-time systems based on dedicated operating systems and standard Linux/Windows-based operating systems. In addition, there are networks of varying degrees of importance that are logically separated. There is a need to consider and examine the adoption of existing solutions for these systems and networks, such as monitoring products, firewalls, and antivirus products.
\end{itemize}
\item \textbf{Future Research Directions}
To expand the taxonomy and make it more accessible and usable, we propose further research in the following areas.
\begin{itemize}
\item \textbf{Cyber Attacks on Autonomous Systems}
Some aircraft systems are autonomous systems that are designed to respond when needed, without the explicit involvement of the pilot. An example of this is the maneuvering characteristics augmentation system (MCAS)~\cite{mako2020evaluation}, which was designed to stabilize the aircraft without any intervention by the pilot. The MCAS is known to be involved in several air accidents that took place in recent years~\cite{MCAS}.
These systems have the potential to cause signficant damage to an aircraft and its passengers, as the crew's degree of impact on these systems is limited. Research is needed to examine how cyber attacks can affect these systems.
\item \textbf{Human Factor}
Addressing the human factor by training and preparing the aircrew to handle and manage various events is  important in the implementation of defense mechanisms and methods in emergencies.
When mapping the various risks and threats, it is necessary to examine how new and existing security solutions can be adapted based on the skills and knowledge of the aircrew.
Some work dealing with human responses in emergencies has been done; a study conducted by Ivan \textit{et al.}~\cite{smith2020view} examined pilots' responses to various cyber attacks.
Future research must address the need to generate and provide valuable and critical information to the pilot and crew, e.g., data from intrusion detection systems. This can be done by developing new systems, adapting relevant systems, and making the information from such systems clear to enable the crew to handle exceptional events.
 An example of dealing with the clarity of information and its accessibility to the pilot was presented by Habler \textit{et al.}~\cite{habler2021analyzing} who provided a deep learning based solution for the detection of anomalous traffic conditions. 
 The novelty was due to the fact that the study dealt with the need to provide a clear indication to the pilot by using an explainability technique designed to formulate and conveniently present computational model decisions.
\item \textbf{Threat Intelligence Platform}
In order to share knowledge, provide up-to-date examples of threats, collect relevant data, and enrich the taxonomy, there is a need for a uniform threat intelligence platform (TIP) which is accessible to airlines, airports, and manufacturers.
Such a platform will collect, aggregate, and organize threat intelligence data of various formats from multiple sources. The TIP will allow security and threat intelligence teams to easily share threat intelligence data with other stakeholders and security systems.
\end{itemize}
\end{enumerate}
\section{\label{sec:Conclusion}Conclusion and Discussion}
In this paper, we provided a comprehensive overview of aircraft systems and components and their various networks, emphasizing the cyber threats they are exposed to and the impact of a cyber attack on an aircraft's essential capabilities.
We also presented a comprehensive and in-depth taxonomy that standardizes the knowledge on and understanding of known threats to avionics systems identified by industry and academia.

The proposed taxonomy deals with the various stages of cyber attacks. It covers avionics systems' critical infrastructure, including air-ground communication, radio navigation aids, aeronautical surveillance, and system-wide information management. 
Additionally, we addressed the different domains of e-enabled aircraft and provided an analysis of the domains' deployment, emphasizing the points at which various networks overlap.

Based on our review and the issues raised in this paper, we conclude that a taxonomy dedicated to avionics is needed given  the diverse and intricate connections, systems, components, and attack surfaces in avionics systems. There will be many challenges in developing and implementing the required defense systems.
More importantly, though, our research points to the need for comprehensive defensive measures aimed at protecting passengers and crew members and ensuring the safety of the aviation sector.
once developed, these mechanisms can be added to the taxonomy, enriching it as they improve safety.

To accomplish this, results published by security researchers must be verified, more studies dealing with avionics defense systems and mechanisms need to be performed, and a unified threat intelligence platform must be adopted for the benefit of all stakeholders.

\clearpage
\bibliographystyle{ACM-Reference-Format}
\bibliography{main}


\begin{thebibliography}{71}

%%% the Web address.

\ifx \showCODEN    \undefined \def \showCODEN     #1{\unskip}     \fi
\ifx \showDOI      \undefined \def \showDOI       #1{#1}\fi
\ifx \showISBNx    \undefined \def \showISBNx     #1{\unskip}     \fi
\ifx \showISBNxiii \undefined \def \showISBNxiii  #1{\unskip}     \fi
\ifx \showISSN     \undefined \def \showISSN      #1{\unskip}     \fi
\ifx \showLCCN     \undefined \def \showLCCN      #1{\unskip}     \fi
\ifx \shownote     \undefined \def \shownote      #1{#1}          \fi
\ifx \showarticletitle \undefined \def \showarticletitle #1{#1}   \fi
\ifx \showURL      \undefined \def \showURL       {\relax}        \fi
% invisible to TeX
\providecommand\bibfield[2]{#2}
\providecommand\bibinfo[2]{#2}
\providecommand\natexlab[1]{#1}
\providecommand\showeprint[2][]{arXiv:#2}

\bibitem[IFE(2017)]%
        {IFECsystem}
 \bibinfo{year}{2017}\natexlab{}.
\newblock \bibinfo{booktitle}{\emph{how-secure-are-ifec-systems}}.
\newblock
\urldef\tempurl%
\url{http://interactive.aviationtoday.com/how-secure-are-ifec-systems/}
\showURL{%
Retrieved Jun 12, 2022 from \tempurl}


\bibitem[AEEC(2022)]%
        {AEEC}
AEEC \bibinfo{year}{2022}\natexlab{}.
\newblock \bibinfo{booktitle}{\emph{The Airlines Electronics and Engineering
  Committee}}.
\newblock
\urldef\tempurl%
\url{https://www.aviation-ia.com/activities/aeec}
\showURL{%
Retrieved Jul 05, 2022 from \tempurl}


\bibitem[Aerospace Village(2020)]%
        {Loadbles2}
Aerospace Village \bibinfo{year}{2020}\natexlab{}.
\newblock \bibinfo{booktitle}{\emph{DEF CON 28 Aerospace Village: Attacking
  Flight Management Systems}}.
\newblock
\urldef\tempurl%
\url{https://www.youtube.com/watch?v=G4dDRXBikvA}
\showURL{%
Retrieved Feb 12, 2022 from \tempurl}


\bibitem[ARINC664(2005)]%
        {ARINC664}
ARINC664 \bibinfo{year}{2005}\natexlab{}.
\newblock \bibinfo{booktitle}{\emph{ARINC664 AFDX data transmission system for
  aircraft, airbus patent}}.
\newblock
\urldef\tempurl%
\url{https://worldwide.espacenet.com/patent/search?q=pn%3DUS6925088}
\showURL{%
Retrieved Jun 06, 2022 from \tempurl}


\bibitem[ARINC821(2008)]%
        {ARINC821}
ARINC821 \bibinfo{year}{2008}\natexlab{}.
\newblock \bibinfo{booktitle}{\emph{ARINC821 Report}}.
\newblock
\urldef\tempurl%
\url{https://www.aviation-ia.com/products/821-aircraft-network-server-system-nss-functional-definition-2}
\showURL{%
Retrieved April 16, 2022 from \tempurl}


\bibitem[Blog(2021)]%
        {CheckPoint}
\bibfield{author}{\bibinfo{person}{Check~Point Blog}.}
  \bibinfo{year}{2021}\natexlab{}.
\newblock \bibinfo{booktitle}{\emph{Checkpoint's cyber security report 2021}}.
\newblock
\urldef\tempurl%
\url{https://blog.checkpoint.com/2021/06/14/ransomware-attacks-continue-to-surge-hitting-a-93-increase-year-over-year/}
\showURL{%
Retrieved Feb 23, 2022 from \tempurl}


\bibitem[Choudhary et~al\mbox{.}(2022)]%
        {choudhary2022aviation}
\bibfield{author}{\bibinfo{person}{Gaurav Choudhary}, \bibinfo{person}{Vikas
  Sihag}, \bibinfo{person}{Shristi Gupta}, {and} \bibinfo{person}{Shishir~Kumar
  Shandilya}.} \bibinfo{year}{2022}\natexlab{}.
\newblock \showarticletitle{Aviation attacks based on ILS and VOR
  vulnerabilities}.
\newblock  (\bibinfo{year}{2022}).
\newblock


\bibitem[Company(2007)]%
        {BoeingPatent}
\bibfield{author}{\bibinfo{person}{The~Boeing Company}.}
  \bibinfo{year}{2007}\natexlab{}.
\newblock \bibinfo{title}{Boeing e-Enablement patent}.
\newblock
\newblock
\newblock
\shownote{Patent can be found at
  \url{https://patentimages.storage.googleapis.com/9c/12/93/43a3858b71b5aa/WO2007117285A2.pdf}.}.


\bibitem[Costin and Francillon(2012)]%
        {costin2012ghost}
\bibfield{author}{\bibinfo{person}{Andrei Costin} {and}
  \bibinfo{person}{Aur{\'e}lien Francillon}.} \bibinfo{year}{2012}\natexlab{}.
\newblock \showarticletitle{Ghost in the Air (Traffic): On insecurity of ADS-B
  protocol and practical attacks on ADS-B devices}.
\newblock \bibinfo{journal}{\emph{Black Hat USA}} (\bibinfo{year}{2012}),
  \bibinfo{pages}{1--12}.
\newblock


\bibitem[Crash Report(2019)]%
        {MCAS}
Crash Report \bibinfo{year}{2019}\natexlab{}.
\newblock \bibinfo{booktitle}{\emph{Preliminary Crash Report}}.
\newblock
\urldef\tempurl%
\url{https://www.npr.org/2019/04/04/709766379/preliminary-crash-report-says-ethiopian-airlines-crew-complied-with-procedures}
\showURL{%
Retrieved Feb 10, 2022 from \tempurl}


\bibitem[Dave et~al\mbox{.}(2022)]%
        {dave2022cyber}
\bibfield{author}{\bibinfo{person}{Gaurav Dave}, \bibinfo{person}{Gaurav
  Choudhary}, \bibinfo{person}{Vikas Sihag}, \bibinfo{person}{Ilsun You}, {and}
  \bibinfo{person}{Kim-Kwang~Raymond Choo}.} \bibinfo{year}{2022}\natexlab{}.
\newblock \showarticletitle{Cyber security challenges in aviation
  communication, navigation, and surveillance}.
\newblock \bibinfo{journal}{\emph{Computers \& Security}}
  \bibinfo{volume}{112} (\bibinfo{year}{2022}), \bibinfo{pages}{102516}.
\newblock


\bibitem[Di~Marco et~al\mbox{.}(2016)]%
        {di2016security}
\bibfield{author}{\bibinfo{person}{Doris Di~Marco}, \bibinfo{person}{Alessandro
  Manzo}, \bibinfo{person}{Marco Ivaldi}, {and} \bibinfo{person}{John Hird}.}
  \bibinfo{year}{2016}\natexlab{}.
\newblock \showarticletitle{Security testing with controller-pilot data link
  communications}. In \bibinfo{booktitle}{\emph{2016 11th International
  Conference on Availability, Reliability and Security (ARES)}}. IEEE,
  \bibinfo{pages}{526--531}.
\newblock


\bibitem[Duchamp et~al\mbox{.}(2016)]%
        {duchamp2016cyber}
\bibfield{author}{\bibinfo{person}{H{\'e}l{\`e}ne Duchamp},
  \bibinfo{person}{Ibrahim Bayram}, {and} \bibinfo{person}{Ranim Korhani}.}
  \bibinfo{year}{2016}\natexlab{}.
\newblock \showarticletitle{Cyber-Security, a new challenge for the aviation
  and automotive industries}. In \bibinfo{booktitle}{\emph{Seminar in
  Information Systems: Applied Cybersecurity Strategy for Managers}}.
  \bibinfo{pages}{1--4}.
\newblock


\bibitem[Elmarady and Rahouma(2021)]%
        {elmarady2021studying}
\bibfield{author}{\bibinfo{person}{Ahmed~Abdelwahab Elmarady} {and}
  \bibinfo{person}{Kamel Rahouma}.} \bibinfo{year}{2021}\natexlab{}.
\newblock \showarticletitle{Studying Cybersecurity in Civil Aviation, Including
  Developing and Applying Aviation Cybersecurity Risk Assessment}.
\newblock \bibinfo{journal}{\emph{IEEE Access}}  \bibinfo{volume}{9}
  (\bibinfo{year}{2021}), \bibinfo{pages}{143997--144016}.
\newblock


\bibitem[Eskilsson et~al\mbox{.}(2020)]%
        {eskilsson2020demonstrating}
\bibfield{author}{\bibinfo{person}{Sofie Eskilsson}, \bibinfo{person}{Hanna
  Gustafsson}, \bibinfo{person}{Suleman Khan}, {and} \bibinfo{person}{Andrei
  Gurtov}.} \bibinfo{year}{2020}\natexlab{}.
\newblock \showarticletitle{Demonstrating ADS-B and CPDLC Attacks with
  Software-Defined Radio}. In \bibinfo{booktitle}{\emph{2020 Integrated
  Communications Navigation and Surveillance Conference (ICNS)}}. IEEE,
  \bibinfo{pages}{1B2--1}.
\newblock


\bibitem[FAA DB(2022)]%
        {FAADATABASE}
FAA DB \bibinfo{year}{2022}\natexlab{}.
\newblock \bibinfo{booktitle}{\emph{FAA Data Base}}.
\newblock
\urldef\tempurl%
\url{https://www.faa.gov/airports/engineering/aircraft_char_database}
\showURL{%
Retrieved Jun 07, 2022 from \tempurl}


\bibitem[Feh{\'e}r and Sandor(2018)]%
        {feher2018effects}
\bibfield{author}{\bibinfo{person}{D{\'a}vid~J{\'a}nos Feh{\'e}r} {and}
  \bibinfo{person}{Barnab{\'a}s Sandor}.} \bibinfo{year}{2018}\natexlab{}.
\newblock \showarticletitle{Effects of the WPA2 KRACK attack in real
  environment}. In \bibinfo{booktitle}{\emph{2018 IEEE 16th international
  symposium on intelligent systems and informatics (SISY)}}. IEEE,
  \bibinfo{pages}{000239--000242}.
\newblock


\bibitem[Gibbs(2015)]%
        {ChrisRoberts}
\bibfield{author}{\bibinfo{person}{Samuel Gibbs}.}
  \bibinfo{year}{2015}\natexlab{}.
\newblock \bibinfo{booktitle}{\emph{Aviation experts dispute hacker’s claim
  he seized control of airliner mid-flight}}.
\newblock
\urldef\tempurl%
\url{https://www.theguardian.com/technology/2015/may/19/hacker-chris-roberts-claim-seized-control-boeing-airliner-disputed-experts}
\showURL{%
Retrieved May 13, 2022 from \tempurl}


\bibitem[Graziano(2021)]%
        {graziano2021establishment}
\bibfield{author}{\bibinfo{person}{Timothy~Michael Graziano}.}
  \bibinfo{year}{2021}\natexlab{}.
\newblock \emph{\bibinfo{title}{Establishment of a Cyber-Physical Systems (CPS)
  Test Bed to Explore Traffic Collision Avoidance System (TCAS) Vulnerabilities
  to Cyber Attacks}}.
\newblock \bibinfo{thesistype}{Ph.\,D. Dissertation}. \bibinfo{school}{Virginia
  Tech}.
\newblock


\bibitem[Gurtov et~al\mbox{.}(2018)]%
        {gurtov2018controller}
\bibfield{author}{\bibinfo{person}{Andrei Gurtov}, \bibinfo{person}{Tatiana
  Polishchuk}, {and} \bibinfo{person}{Max Wernberg}.}
  \bibinfo{year}{2018}\natexlab{}.
\newblock \showarticletitle{Controller--pilot data link communication
  security}.
\newblock \bibinfo{journal}{\emph{Sensors}} \bibinfo{volume}{18},
  \bibinfo{number}{5} (\bibinfo{year}{2018}), \bibinfo{pages}{1636}.
\newblock


\bibitem[Habler and Shabtai(2021)]%
        {habler2021analyzing}
\bibfield{author}{\bibinfo{person}{Edan Habler} {and} \bibinfo{person}{Asaf
  Shabtai}.} \bibinfo{year}{2021}\natexlab{}.
\newblock \showarticletitle{Analyzing Sequences of Airspace States to Detect
  Anomalous Traffic Conditions}.
\newblock \bibinfo{journal}{\emph{IEEE Trans. Aerospace Electron. Systems}}
  (\bibinfo{year}{2021}).
\newblock


\bibitem[Hagm{\"u}ller et~al\mbox{.}(2004)]%
        {hagmuller2004speech}
\bibfield{author}{\bibinfo{person}{Martin Hagm{\"u}ller},
  \bibinfo{person}{Horst Hering}, \bibinfo{person}{Andreas Kr{\"o}pfl}, {and}
  \bibinfo{person}{Gernot Kubin}.} \bibinfo{year}{2004}\natexlab{}.
\newblock \showarticletitle{Speech watermarking for air traffic control}. In
  \bibinfo{booktitle}{\emph{2004 12th European Signal Processing Conference}}.
  IEEE, \bibinfo{pages}{1653--1656}.
\newblock


\bibitem[Hannah et~al\mbox{.}(2020)]%
        {hannah2020traffic}
\bibfield{author}{\bibinfo{person}{John Hannah}, \bibinfo{person}{Robert
  Mills}, {and} \bibinfo{person}{Richard Dill}.}
  \bibinfo{year}{2020}\natexlab{}.
\newblock \showarticletitle{Traffic collision avoidance system: threat actor
  model and attack taxonomy}. In \bibinfo{booktitle}{\emph{2020 New Trends in
  Civil Aviation (NTCA)}}. IEEE, \bibinfo{pages}{17--26}.
\newblock


\bibitem[Hannah(2021)]%
        {hannah2021cyber}
\bibfield{author}{\bibinfo{person}{John~W Hannah}.}
  \bibinfo{year}{2021}\natexlab{}.
\newblock \showarticletitle{A Cyber Threat Taxonomy and a Viability Analysis
  for False Injections in the TCAS}.
\newblock  (\bibinfo{year}{2021}).
\newblock


\bibitem[Hering et~al\mbox{.}(2003)]%
        {hering2003safety}
\bibfield{author}{\bibinfo{person}{Horst Hering}, \bibinfo{person}{G Kubin},
  {et~al\mbox{.}}} \bibinfo{year}{2003}\natexlab{}.
\newblock \showarticletitle{Safety and security increase for air traffic
  management through unnoticeable watermark aircraft identification tag
  transmitted with the VHF voice communication}. In
  \bibinfo{booktitle}{\emph{Digital Avionics Systems Conference, 2003. DASC'03.
  The 22nd}}, Vol.~\bibinfo{volume}{1}. IEEE, \bibinfo{pages}{4--E}.
\newblock


\bibitem[Hutchins et~al\mbox{.}(2011)]%
        {hutchins2011intelligence}
\bibfield{author}{\bibinfo{person}{Eric~M Hutchins}, \bibinfo{person}{Michael~J
  Cloppert}, \bibinfo{person}{Rohan~M Amin}, {et~al\mbox{.}}}
  \bibinfo{year}{2011}\natexlab{}.
\newblock \showarticletitle{Intelligence-driven computer network defense
  informed by analysis of adversary campaigns and intrusion kill chains}.
\newblock \bibinfo{journal}{\emph{Leading Issues in Information Warfare \&
  Security Research}} \bibinfo{volume}{1}, \bibinfo{number}{1}
  (\bibinfo{year}{2011}), \bibinfo{pages}{80}.
\newblock


\bibitem[IATA Best Practices(2020)]%
        {IATALOADABLE}
IATA Best Practices \bibinfo{year}{2020}\natexlab{}.
\newblock \bibinfo{booktitle}{\emph{IATA - Best Practices for Loadable Software
  Management and Configuration}}.
\newblock
\urldef\tempurl%
\url{https://docplayer.net/15768767-Best-practices-for-loadable-software-management-and-configuration-control.html}
\showURL{%
Retrieved Jan 08, 2022 from \tempurl}


\bibitem[IOActive(2016)]%
        {IFECIOACTIVE}
\bibfield{author}{\bibinfo{person}{IOActive}.} \bibinfo{year}{2016}\natexlab{}.
\newblock
  \bibinfo{booktitle}{\emph{ioactive-discovers-in-flight-entertainment-system-vulnerabilities}}.
\newblock
\urldef\tempurl%
\url{https://ioactive.com/article/ioactive-discovers-in-flight-entertainment-system-vulnerabilities/}
\showURL{%
Retrieved Feb 11, 2022 from \tempurl}


\bibitem[Kirby(2014)]%
        {EFBReport}
\bibfield{author}{\bibinfo{person}{Mary Kirby}.}
  \bibinfo{year}{2014}\natexlab{}.
\newblock \bibinfo{booktitle}{\emph{Most airlines lack EFB cyber-security plan:
  report}}.
\newblock \bibinfo{type}{{T}echnical {R}eport}.
\newblock


\bibitem[Kovzovic and Durdevic(2021)]%
        {kovzovic2021spoofing}
\bibfield{author}{\bibinfo{person}{Dejan~V Kovzovic} {and}
  \bibinfo{person}{Dragan~Z Durdevic}.} \bibinfo{year}{2021}\natexlab{}.
\newblock \showarticletitle{Spoofing in aviation: Security threats on GPS and
  ADS-B systems}.
\newblock \bibinfo{journal}{\emph{Vojnotehnicki glasnik/Military Technical
  Courier}} \bibinfo{volume}{69}, \bibinfo{number}{2} (\bibinfo{year}{2021}),
  \bibinfo{pages}{461--485}.
\newblock


\bibitem[Lykou et~al\mbox{.}(2019)]%
        {lykou2019aviation}
\bibfield{author}{\bibinfo{person}{Georgia Lykou}, \bibinfo{person}{George
  Iakovakis}, {and} \bibinfo{person}{Dimitris Gritzalis}.}
  \bibinfo{year}{2019}\natexlab{}.
\newblock \showarticletitle{Aviation cybersecurity and cyber-resilience:
  assessing risk in air traffic management}.
\newblock In \bibinfo{booktitle}{\emph{Critical Infrastructure Security and
  Resilience}}. \bibinfo{publisher}{Springer}, \bibinfo{pages}{245--260}.
\newblock


\bibitem[Mako et~al\mbox{.}(2020)]%
        {mako2020evaluation}
\bibfield{author}{\bibinfo{person}{Sebasti{\'a}n Mako}, \bibinfo{person}{Marek
  Pilat}, \bibinfo{person}{P Svab}, \bibinfo{person}{J Kozuba}, {and}
  \bibinfo{person}{M Cicvakova}.} \bibinfo{year}{2020}\natexlab{}.
\newblock \showarticletitle{Evaluation of MCAS system}.
\newblock \bibinfo{journal}{\emph{Acta Avionica J.}}  \bibinfo{volume}{40}
  (\bibinfo{year}{2020}), \bibinfo{pages}{21--28}.
\newblock


\bibitem[Miralles et~al\mbox{.}(2020)]%
        {miralles2020assessment}
\bibfield{author}{\bibinfo{person}{Damian Miralles}, \bibinfo{person}{Aurelie
  Bornot}, \bibinfo{person}{Paul Rouquette}, \bibinfo{person}{Nathan Levigne},
  \bibinfo{person}{Dennis~M Akos}, \bibinfo{person}{Yu-Hsuan Chen},
  \bibinfo{person}{Sherman Lo}, {and} \bibinfo{person}{Todd Walter}.}
  \bibinfo{year}{2020}\natexlab{}.
\newblock \showarticletitle{An assessment of GPS spoofing detection via radio
  power and signal quality monitoring for aviation safety operations}.
\newblock \bibinfo{journal}{\emph{IEEE Intelligent Transportation Systems
  Magazine}} \bibinfo{volume}{12}, \bibinfo{number}{3} (\bibinfo{year}{2020}),
  \bibinfo{pages}{136--146}.
\newblock


\bibitem[Mirzaei et~al\mbox{.}(2019)]%
        {mirzaei2019security}
\bibfield{author}{\bibinfo{person}{Kayvan~Faghih Mirzaei},
  \bibinfo{person}{Bruno~Pessanha de Carvalho}, {and} \bibinfo{person}{Patrick
  Pschorn}.} \bibinfo{year}{2019}\natexlab{}.
\newblock \bibinfo{booktitle}{\emph{Security of ADS-B: Attack Scenarios}}.
\newblock \bibinfo{type}{{T}echnical {R}eport}.
  \bibinfo{institution}{EasyChair}.
\newblock


\bibitem[MITRE Website(2022)]%
        {MITRE}
MITRE Website \bibinfo{year}{2022}\natexlab{}.
\newblock \bibinfo{booktitle}{\emph{MITRE ATT\&CK}}.
\newblock
\urldef\tempurl%
\url{https://attack.mitre.org/}
\showURL{%
Retrieved Jun 17, 2022 from \tempurl}


\bibitem[Moser et~al\mbox{.}(2016)]%
        {moser2016investigation}
\bibfield{author}{\bibinfo{person}{Daniel Moser}, \bibinfo{person}{Patrick
  Leu}, \bibinfo{person}{Vincent Lenders}, \bibinfo{person}{Aanjhan
  Ranganathan}, \bibinfo{person}{Fabio Ricciato}, {and} \bibinfo{person}{Srdjan
  Capkun}.} \bibinfo{year}{2016}\natexlab{}.
\newblock \showarticletitle{Investigation of multi-device location spoofing
  attacks on air traffic control and possible countermeasures}. In
  \bibinfo{booktitle}{\emph{Proceedings of the 22nd Annual International
  Conference on Mobile Computing and Networking}}. \bibinfo{pages}{375--386}.
\newblock


\bibitem[Mostafa et~al\mbox{.}(2016)]%
        {mostafa2016vulnerability}
\bibfield{author}{\bibinfo{person}{Mohamad Mostafa}, \bibinfo{person}{Okuary
  Osechas}, {and} \bibinfo{person}{Michael Schnell}.}
  \bibinfo{year}{2016}\natexlab{}.
\newblock \showarticletitle{Vulnerability analysis of the CNS-infrastructure:
  An exemplarily approach}. In \bibinfo{booktitle}{\emph{2016 IEEE/AIAA 35th
  Digital Avionics Systems Conference (DASC)}}. IEEE, \bibinfo{pages}{1--9}.
\newblock


\bibitem[Ochieng et~al\mbox{.}(2003)]%
        {ochieng2003gps}
\bibfield{author}{\bibinfo{person}{Washington~Y Ochieng}, \bibinfo{person}{Knut
  Sauer}, \bibinfo{person}{David Walsh}, \bibinfo{person}{Gary Brodin},
  \bibinfo{person}{Steve Griffin}, {and} \bibinfo{person}{Mark Denney}.}
  \bibinfo{year}{2003}\natexlab{}.
\newblock \showarticletitle{GPS integrity and potential impact on aviation
  safety}.
\newblock \bibinfo{journal}{\emph{The journal of navigation}}
  \bibinfo{volume}{56}, \bibinfo{number}{1} (\bibinfo{year}{2003}),
  \bibinfo{pages}{51--65}.
\newblock


\bibitem[Osechas et~al\mbox{.}(2017)]%
        {osechas2017addressing}
\bibfield{author}{\bibinfo{person}{Okuary Osechas}, \bibinfo{person}{Mohamad
  Mostafa}, \bibinfo{person}{Thomas Graupl}, {and} \bibinfo{person}{Michael
  Meurer}.} \bibinfo{year}{2017}\natexlab{}.
\newblock \showarticletitle{Addressing vulnerabilities of the CNS
  infrastructure to targeted radio interference}.
\newblock \bibinfo{journal}{\emph{IEEE Aerospace and Electronic Systems
  Magazine}} \bibinfo{volume}{32}, \bibinfo{number}{11} (\bibinfo{year}{2017}),
  \bibinfo{pages}{34--42}.
\newblock


\bibitem[Partners(2021)]%
        {EFBPPT}
\bibfield{author}{\bibinfo{person}{Pen~Test Partners}.}
  \bibinfo{year}{2021}\natexlab{}.
\newblock \bibinfo{booktitle}{\emph{EFB-tampering}}.
\newblock \bibinfo{type}{{T}echnical {R}eport}.
\newblock


\bibitem[PenTestPartners(2020a)]%
        {ILSandTCASSpooging}
\bibfield{author}{\bibinfo{person}{PenTestPartners}.}
  \bibinfo{year}{2020}\natexlab{a}.
\newblock \bibinfo{booktitle}{\emph{DEF CON 28: ILS and TCAS Spoofing}}.
\newblock
\urldef\tempurl%
\url{https://www.pentestpartners.com/security-blog/ils-and-tcas-spoofing/}
\showURL{%
Retrieved May 20, 2022 from \tempurl}


\bibitem[PenTestPartners(2020b)]%
        {IntroToAcars}
\bibfield{author}{\bibinfo{person}{PenTestPartners}.}
  \bibinfo{year}{2020}\natexlab{b}.
\newblock \bibinfo{booktitle}{\emph{DEF CON 28: Introduction to Acars}}.
\newblock
\urldef\tempurl%
\url{https://www.pentestpartners.com/security-blog/introduction-to-acars}
\showURL{%
Retrieved May 20, 2022 from \tempurl}


\bibitem[Pollack and Ranganathan(2018)]%
        {pollack2018aviation}
\bibfield{author}{\bibinfo{person}{Jason Pollack} {and}
  \bibinfo{person}{Prakash Ranganathan}.} \bibinfo{year}{2018}\natexlab{}.
\newblock \showarticletitle{Aviation navigation systems security: ADS-B, GPS,
  iff}. In \bibinfo{booktitle}{\emph{Proceedings of the International
  Conference on Security and Management (SAM)}}. The Steering Committee of The
  World Congress in Computer Science, Computer~…, \bibinfo{pages}{129--135}.
\newblock


\bibitem[PTsecurity(2018)]%
        {ImpactAssessment}
\bibfield{author}{\bibinfo{person}{PTsecurity}.}
  \bibinfo{year}{2018}\natexlab{}.
\newblock \bibinfo{booktitle}{\emph{Impact Assessment of Cyber security
  Threats}}.
\newblock \bibinfo{type}{{T}echnical {R}eport}.
\newblock


\bibitem[PTsecurity(2021)]%
        {PTsecurityRepport}
\bibfield{author}{\bibinfo{person}{PTsecurity}.}
  \bibinfo{year}{2021}\natexlab{}.
\newblock \bibinfo{booktitle}{\emph{PTsecurity report Q1 2021}}.
\newblock \bibinfo{type}{{T}echnical {R}eport}.
\newblock


\bibitem[Santamarta(2014a)]%
        {santamarta2014satcom}
\bibfield{author}{\bibinfo{person}{Ruben Santamarta}.}
  \bibinfo{year}{2014}\natexlab{a}.
\newblock \showarticletitle{SATCOM terminals: Hacking by air, sea, and land}.
\newblock \bibinfo{journal}{\emph{DEFCON White Paper}} (\bibinfo{year}{2014}).
\newblock


\bibitem[Santamarta(2014b)]%
        {santamarta2014wake}
\bibfield{author}{\bibinfo{person}{Ruben Santamarta}.}
  \bibinfo{year}{2014}\natexlab{b}.
\newblock \showarticletitle{A wake-up call for SATCOM security}.
\newblock \bibinfo{journal}{\emph{Technical White Paper}}
  (\bibinfo{year}{2014}).
\newblock


\bibitem[Santamarta(2018)]%
        {santamarta2018last}
\bibfield{author}{\bibinfo{person}{Ruben Santamarta}.}
  \bibinfo{year}{2018}\natexlab{}.
\newblock \bibinfo{booktitle}{\emph{Last call for SATCOM security}}.
\newblock \bibinfo{publisher}{IOActive}.
\newblock


\bibitem[Sathaye et~al\mbox{.}(2019)]%
        {sathaye2019wireless}
\bibfield{author}{\bibinfo{person}{Harshad Sathaye}, \bibinfo{person}{Domien
  Schepers}, \bibinfo{person}{Aanjhan Ranganathan}, {and}
  \bibinfo{person}{Guevara Noubir}.} \bibinfo{year}{2019}\natexlab{}.
\newblock \showarticletitle{Wireless attacks on aircraft instrument landing
  systems}. In \bibinfo{booktitle}{\emph{28th USENIX Security Symposium (USENIX
  Security 19)}}. \bibinfo{pages}{357--372}.
\newblock


\bibitem[Seri et~al\mbox{.}(2019)]%
        {seri2019critical}
\bibfield{author}{\bibinfo{person}{Ben Seri}, \bibinfo{person}{Gregory
  Vishnepolsky}, {and} \bibinfo{person}{Dor Zusman}.}
  \bibinfo{year}{2019}\natexlab{}.
\newblock \showarticletitle{Critical vulnerabilities to remotely compromise
  VxWorks, the most popular RTOS}.
\newblock \bibinfo{journal}{\emph{White Paper, ARMIS, URGENT/11}}
  (\bibinfo{year}{2019}).
\newblock


\bibitem[Shaikh et~al\mbox{.}(2019)]%
        {shaikh2019review}
\bibfield{author}{\bibinfo{person}{Farooq Shaikh}, \bibinfo{person}{Mohamed
  Rahouti}, \bibinfo{person}{Nasir Ghani}, \bibinfo{person}{Kaiqi Xiong},
  \bibinfo{person}{Elias Bou-Harb}, {and} \bibinfo{person}{Jamal Haque}.}
  \bibinfo{year}{2019}\natexlab{}.
\newblock \showarticletitle{A review of recent advances and security challenges
  in emerging E-enabled aircraft systems}.
\newblock \bibinfo{journal}{\emph{IEEE access}}  \bibinfo{volume}{7}
  (\bibinfo{year}{2019}), \bibinfo{pages}{63164--63180}.
\newblock


\bibitem[Shatilin(2015)]%
        {EICASKaspersky}
\bibfield{author}{\bibinfo{person}{Ilja Shatilin}.}
  \bibinfo{year}{2015}\natexlab{}.
\newblock \bibinfo{booktitle}{\emph{Hacking an aircraft: is it already real}}.
\newblock \bibinfo{type}{{T}echnical {R}eport}.
  \bibinfo{institution}{Kaspersky}.
\newblock


\bibitem[Shostack et~al\mbox{.}(2006)]%
        {shostack2006uncover}
\bibfield{author}{\bibinfo{person}{A Shostack}, \bibinfo{person}{S Lambert},
  {and} \bibinfo{person}{S Hernan}.} \bibinfo{year}{2006}\natexlab{}.
\newblock \showarticletitle{Uncover Security Design Flaws using STRIDE}.
\newblock \bibinfo{journal}{\emph{MSDN magazine, November}}
  (\bibinfo{year}{2006}).
\newblock


\bibitem[Smailes et~al\mbox{.}(2021)]%
        {smailes2021you}
\bibfield{author}{\bibinfo{person}{Joshua Smailes}, \bibinfo{person}{Daniel
  Moser}, \bibinfo{person}{Matthew Smith}, \bibinfo{person}{Martin Strohmeier},
  \bibinfo{person}{Vincent Lenders}, {and} \bibinfo{person}{Ivan Martinovic}.}
  \bibinfo{year}{2021}\natexlab{}.
\newblock \showarticletitle{You talkin’to me? Exploring Practical Attacks on
  Controller Pilot Data Link Communications}. In
  \bibinfo{booktitle}{\emph{Proceedings of the 7th ACM on Cyber-Physical System
  Security Workshop}}. \bibinfo{pages}{53--64}.
\newblock


\bibitem[Smith et~al\mbox{.}(2017)]%
        {smith2017analyzing}
\bibfield{author}{\bibinfo{person}{Matthew Smith}, \bibinfo{person}{Daniel
  Moser}, \bibinfo{person}{Martin Strohmeier}, \bibinfo{person}{Vincent
  Lenders}, {and} \bibinfo{person}{Ivan Martinovic}.}
  \bibinfo{year}{2017}\natexlab{}.
\newblock \showarticletitle{Analyzing privacy breaches in the aircraft
  communications addressing and reporting system (acars)}.
\newblock \bibinfo{journal}{\emph{arXiv preprint arXiv:1705.07065}}
  (\bibinfo{year}{2017}).
\newblock


\bibitem[Smith et~al\mbox{.}(2020a)]%
        {smith2020view}
\bibfield{author}{\bibinfo{person}{Matthew Smith}, \bibinfo{person}{Martin
  Strohmeier}, \bibinfo{person}{Jon Harman}, \bibinfo{person}{Vincent Lenders},
  {and} \bibinfo{person}{Ivan Martinovic}.} \bibinfo{year}{2020}\natexlab{a}.
\newblock \showarticletitle{A view from the cockpit: exploring pilot reactions
  to attacks on avionic systems}.
\newblock  (\bibinfo{year}{2020}).
\newblock


\bibitem[Smith et~al\mbox{.}(2016)]%
        {smith2016security}
\bibfield{author}{\bibinfo{person}{Matt Smith}, \bibinfo{person}{Martin
  Strohmeier}, \bibinfo{person}{Vincent Lenders}, {and} \bibinfo{person}{Ivan
  Martinovic}.} \bibinfo{year}{2016}\natexlab{}.
\newblock \showarticletitle{On the security and privacy of ACARS}. In
  \bibinfo{booktitle}{\emph{2016 Integrated Communications Navigation and
  Surveillance (ICNS)}}. IEEE, \bibinfo{pages}{1--27}.
\newblock


\bibitem[Smith et~al\mbox{.}(2020b)]%
        {smith2020understanding}
\bibfield{author}{\bibinfo{person}{Matthew Smith}, \bibinfo{person}{Martin
  Strohmeier}, \bibinfo{person}{Vincent Lenders}, {and} \bibinfo{person}{Ivan
  Martinovic}.} \bibinfo{year}{2020}\natexlab{b}.
\newblock \showarticletitle{Understanding Realistic Attacks on Airborne
  Collision Avoidance Systems}.
\newblock \bibinfo{journal}{\emph{arXiv preprint arXiv:2010.01034}}
  (\bibinfo{year}{2020}).
\newblock


\bibitem[Specifications(2022)]%
        {specifications}
Specifications \bibinfo{year}{2022}\natexlab{}.
\newblock \bibinfo{booktitle}{\emph{Aircraft specifications}}.
\newblock
\urldef\tempurl%
\url{http://www.axonaviation.com/commercial-aircraft/aircraft-data/aircraft-specifications}
\showURL{%
Retrieved May 01, 2022 from \tempurl}


\bibitem[STRIDE(2009)]%
        {STRIDEMODEL}
STRIDE \bibinfo{year}{2009}\natexlab{}.
\newblock \bibinfo{booktitle}{\emph{STRIDE Threat Model}}.
\newblock
\urldef\tempurl%
\url{https://docs.microsoft.com/en-us/previous-versions/commerce-server/ee823878(v=cs.20)?redirectedfrom=MSDN}
\showURL{%
Retrieved November 12, 2009 from \tempurl}


\bibitem[Strohmeier et~al\mbox{.}(2020)]%
        {strohmeier2020securing}
\bibfield{author}{\bibinfo{person}{Martin Strohmeier}, \bibinfo{person}{Ivan
  Martinovic}, {and} \bibinfo{person}{Vincent Lenders}.}
  \bibinfo{year}{2020}\natexlab{}.
\newblock \showarticletitle{Securing the air--ground link in aviation}.
\newblock In \bibinfo{booktitle}{\emph{The Security of Critical
  Infrastructures}}. \bibinfo{publisher}{Springer}, \bibinfo{pages}{131--154}.
\newblock


\bibitem[Strohmeier et~al\mbox{.}(2016)]%
        {strohmeier2016perception}
\bibfield{author}{\bibinfo{person}{Martin Strohmeier},
  \bibinfo{person}{Matthias Sch{\"a}fer}, \bibinfo{person}{Rui Pinheiro},
  \bibinfo{person}{Vincent Lenders}, {and} \bibinfo{person}{Ivan Martinovic}.}
  \bibinfo{year}{2016}\natexlab{}.
\newblock \showarticletitle{On perception and reality in wireless air traffic
  communication security}.
\newblock \bibinfo{journal}{\emph{IEEE transactions on intelligent
  transportation systems}} \bibinfo{volume}{18}, \bibinfo{number}{6}
  (\bibinfo{year}{2016}), \bibinfo{pages}{1338--1357}.
\newblock


\bibitem[Strom et~al\mbox{.}(2018)]%
        {strom2018mitre}
\bibfield{author}{\bibinfo{person}{Blake~E Strom}, \bibinfo{person}{Andy
  Applebaum}, \bibinfo{person}{Doug~P Miller}, \bibinfo{person}{Kathryn~C
  Nickels}, \bibinfo{person}{Adam~G Pennington}, {and} \bibinfo{person}{Cody~B
  Thomas}.} \bibinfo{year}{2018}\natexlab{}.
\newblock \showarticletitle{Mitre att\&ck: Design and philosophy}.
\newblock \bibinfo{journal}{\emph{Technical report}} (\bibinfo{year}{2018}).
\newblock


\bibitem[Tanil et~al\mbox{.}(2016)]%
        {tanil2016ins}
\bibfield{author}{\bibinfo{person}{Cagatay Tanil}, \bibinfo{person}{Samer
  Khanafseh}, {and} \bibinfo{person}{Boris Pervan}.}
  \bibinfo{year}{2016}\natexlab{}.
\newblock \showarticletitle{An INS monitor against GNSS spoofing attacks during
  GBAS and SBAS-assisted aircraft landing approaches}. In
  \bibinfo{booktitle}{\emph{Proceedings of the 29th International Technical
  Meeting of The Satellite Division of the Institute of Navigation (ION GNSS+
  2016)}}. \bibinfo{pages}{2981--2990}.
\newblock


\bibitem[Teso(2013a)]%
        {teso2013aircraft}
\bibfield{author}{\bibinfo{person}{Hugo Teso}.}
  \bibinfo{year}{2013}\natexlab{a}.
\newblock \showarticletitle{Aircraft hacking: Practical aero series}. In
  \bibinfo{booktitle}{\emph{4th Hack in the Box Security Conference in
  Europe}}.
\newblock


\bibitem[Teso(2013b)]%
        {Loadbles1}
\bibfield{author}{\bibinfo{person}{Hugo Teso}.}
  \bibinfo{year}{2013}\natexlab{b}.
\newblock \bibinfo{booktitle}{\emph{Hugo Teso - Digging Deeper Into Aviation
  Security}}.
\newblock
\urldef\tempurl%
\url{https://conference.hitb.org/hitbsecconf2013kul/materials/D2T1%20-%20Hugo%20Teso%20-%20Digging%20Deeper%20Into%20Aviation%20Security.pdf}
\showURL{%
Retrieved Feb 12, 2022 from \tempurl}


\bibitem[Thomas et~al\mbox{.}(2009)]%
        {thomas2009rise}
\bibfield{author}{\bibinfo{person}{Vinoo Thomas}, \bibinfo{person}{Prashanth
  Ramagopal}, {and} \bibinfo{person}{Rahul Mohandas}.}
  \bibinfo{year}{2009}\natexlab{}.
\newblock \showarticletitle{The rise of autorun-based malware}.
\newblock \bibinfo{journal}{\emph{McAfee Avert Labs., McAfee Inc}}
  (\bibinfo{year}{2009}).
\newblock


\bibitem[Truffer et~al\mbox{.}(2017)]%
        {truffer2017jamming}
\bibfield{author}{\bibinfo{person}{Pascal Truffer}, \bibinfo{person}{Maurizio
  Scaramuzza}, \bibinfo{person}{Marc Troller}, {and} \bibinfo{person}{Marc
  Bertschi}.} \bibinfo{year}{2017}\natexlab{}.
\newblock \showarticletitle{Jamming of aviation GPS receivers: Investigation of
  field trials performed with civil and military aircraft}. In
  \bibinfo{booktitle}{\emph{Proceedings of the 30th International Technical
  Meeting of the Satellite Division of The Institute of Navigation (ION GNSS+
  2017)}}. \bibinfo{pages}{1258--1266}.
\newblock


\bibitem[Vanhoef and Piessens(2017)]%
        {vanhoef2017key}
\bibfield{author}{\bibinfo{person}{Mathy Vanhoef} {and} \bibinfo{person}{Frank
  Piessens}.} \bibinfo{year}{2017}\natexlab{}.
\newblock \showarticletitle{Key reinstallation attacks: Forcing nonce reuse in
  WPA2}. In \bibinfo{booktitle}{\emph{Proceedings of the 2017 ACM SIGSAC
  Conference on Computer and Communications Security}}.
  \bibinfo{pages}{1313--1328}.
\newblock


\bibitem[Viveros(2016)]%
        {viveros2016analysis}
\bibfield{author}{\bibinfo{person}{Camilo Andres~Pantoja Viveros}.}
  \bibinfo{year}{2016}\natexlab{}.
\newblock \emph{\bibinfo{title}{Analysis of the cyber attacks against ADS-B
  perspective of aviation experts}}.
\newblock \bibinfo{thesistype}{Ph.\,D. Dissertation}.
  \bibinfo{school}{University of Tartu Tartu, Estonia}.
\newblock


\bibitem[Zhang et~al\mbox{.}(2017)]%
        {zhang2017analysis}
\bibfield{author}{\bibinfo{person}{Ru Zhang}, \bibinfo{person}{Gongshen Liu},
  \bibinfo{person}{Jianyi Liu}, {and} \bibinfo{person}{Jan~P Nees}.}
  \bibinfo{year}{2017}\natexlab{}.
\newblock \showarticletitle{Analysis of message attacks in aviation data-link
  communication}.
\newblock \bibinfo{journal}{\emph{IEEE Access}}  \bibinfo{volume}{6}
  (\bibinfo{year}{2017}), \bibinfo{pages}{455--463}.
\newblock


\end{thebibliography}
\clearpage

\appendix
\section{Appendices}
\begin{table}[b]
\scriptsize
\centering
\begin{tabular}{|c|c|}
\hline
\textbf{Acronym} & \textbf{Meaning}                                        \\ \hline
ABAS             & Aircraft-based augmentation system                      \\ \hline
ACARS            & Aircraft communications addressing and reporting system \\ \hline
ACD              & Aircraft control domain                                 \\ \hline
ACE              & Actuator control electronics                            \\ \hline
ACMS             & Aircraft condition monitoring system                    \\ \hline
ADN              & Aircraft data network                                   \\ \hline
ADS-B            & Automatic dependent surveillance-broadcast              \\ \hline
AEEC             & Airlines electronic engineering committee               \\ \hline
AISD             & Airline information services domain                     \\ \hline
AMI              & Airline modifiable information                          \\ \hline
AOC              & Aircraft operational control                            \\ \hline
ATC              & Air traffic control                                     \\ \hline
ATM              & Air traffic management                                  \\ \hline
ATN              & Aeronautical telecommunication network                  \\ \hline
CIS-MS           & Crew information system/maintenance system            \\ \hline
CPDLC            & Controller pilot data link communications               \\ \hline
CSM              & Controller server module                                \\ \hline
CSS              & Cabin services system                                   \\ \hline
CWLU             & Connectivity and crew wireless LAN unit                 \\ \hline
DME              & Distance measuring equipment                            \\ \hline 
DSP              & Datalink service provider                               \\ \hline
EASA             & European Union Aviation Safety Agency                   \\ \hline
ECAM             & Electronic centralized aircraft monitor                 \\ \hline
EFB              & Electronic flight bag                                   \\ \hline
EFIS             & Instrument display system                               \\ \hline
EGM              & Ethernet gateway module                                 \\ \hline
EICAS            & Engine-indicating and crew-alerting system              \\ \hline
FANS             & Future air navigation system                            \\ \hline
FDR              & Flight data recorder                                    \\ \hline
FMS              & Flight management systems                               \\ \hline
FSM              & File server module                                      \\ \hline
GBAS             & Ground-based augmentation system                        \\ \hline
GNSS             & Global navigation satellite system                      \\ \hline
GPS              & Global navigation satellite                             \\ \hline
HF               & High frequency                                          \\ \hline
IFE              & In-flight entertainment                                 \\ \hline
IFR              & Instrument flight rules                                 \\ \hline
ILS              & Instrument landing system                               \\ \hline
IMA              & Integrated modular avionics                             \\ \hline
LRU              & Loadable replaceable unit                               \\ \hline
MFD              & Multi-function display                                  \\ \hline
NIM              & Network interface module                                \\ \hline
OPC              & Operational program configuration                       \\ \hline
OPS              & Operational program software                            \\ \hline
PFD              & Primary flight display                                  \\ \hline
PIESD            & Passenger information and entertainment domain          \\ \hline
PODD             & Passenger owned devices domain                          \\ \hline
RDC              & Remote data concentrator                                \\ \hline
SATCOM           & Aircraft satellite communication system                 \\ \hline
SBAS             & Satellite-based augmentation system                     \\ \hline
SDR              & Software-defined radio                                  \\ \hline
SDU              & Satellite data unit                                     \\ \hline
TCAS             & Traffic collision avoidance system                      \\ \hline
TWLU             & Terminal wireless LAN unit                              \\ \hline
UHF              & Ultra high frequency                                    \\ \hline
VFR              & Visual flight rules                                     \\ \hline
VHF              & Very high frequency                                     \\ \hline
\end{tabular}
\caption{Acronym table.}
\label{tab:my-table}
\end{table}

\end{document}